\documentclass[eqsecnum,aps,preprint]{revtex4}
\newcommand{\be}{\begin{eqnarray}&&}
\usepackage{graphicx}
\usepackage{wrapfig}
\font\tenbifull=cmmib10 scaled 1200 % bold math italic
\font\tenbimed=cmmib9
\font\tenbismall=cmmib7
\def\bmit{\fam9 }
\mathchardef\bbkappa="7114
\mathchardef\bbrho="711A
\mathchardef\bbsigma="711B
\mathchardef\bbtau="711C
\mathchardef\bbvarrho="7125
\mathchardef\bbvarsigma="7126
\mathchardef\bbPhi="7008
\mathchardef\bbxi="7118

\def\boldsigma{{\bmit\bbsigma}}
\def\boldtau{{\bmit\bbtau}}

\def\boldPhi{{\bmit\bbPhi}}
\def\boldxi{{\bmit\bbxi}}
\newcommand {\CM} {{\cal M}}
\newcommand{\ee}{\end{eqnarray}}
\def\bn{{\mbox{\boldmath$n$}}}

\begin{document}
\thispagestyle{empty}
\title{Combined Analysis of Near-Threshold Production of
$\mbox{\boldmath$\omega$}$ and  $\mbox{\boldmath$\phi$}$ Mesons in Nucleon-Nucleon Collisions
within an Effective Meson-Nucleon Model}

\author{L.P. Kaptari}
\altaffiliation{On leave of absence from Bogoliubov Lab. Theoretical Physics, JINR, 141980 Dubna,
Russia}
\author{ B. K\"ampfer}
\affiliation{Forschungszentrum Rossendorf, PF 510119, 01314 Dresden, Germany}

%\date{\today}

\begin{abstract}
Vector meson ($V = \omega, \phi$) production in near-threshold elementary 
nucleon-nucleon collisions $pp\to ppV$, $pn\to pnV$ and $pn\to dV$ 
is studied  within an effective meson-nucleon theory. 
It is shown that a set of effective parameters can be established to describe fairly well the
available experimental data of angular distributions and the energy dependence of the total cross
sections without explicit implementation of the Okubo-Zweig-Iisuka rule violation. Isospin
effects are considered in detail and compared with experimental data whenever available.
\end{abstract}
\maketitle

\section{Introduction}

A combined theoretical analysis of $\omega$ and $\phi$ meson production in the processes $pp\to
ppV$, $pn\to pnV$ and $pn\to dV$ (here $V$ denotes a vector meson [$\omega$ or $\phi$], 
$p$ ($n$) denotes a proton (neutron), and 
$d$ stands for the deuteron in the final state) at near-threshold
energies is interesting for different aspects of contemporary  particle and nuclear physics. For
instance, according  to the Okubo-Zweig-Iizuka (OZI) rule~\cite{ozi} the production of $\phi$
mesons in nucleon-nucleon collisions should be strongly suppressed relative to $\omega$
production. An enhanced $\phi$ production  would imply some exotic (e.g., hidden strangeness)
components in the nucleon wave function. The OZI rule is based on Sakurai's observation
\cite{sakurai} that the lowest $1^-$ vector mesons obey the Gell-Mann SU(3) octet classification
\cite{gelmann} and the Gell-Mann--Okubo mass formulae only if one attaches to the eight SU(3)
matrices $\lambda_{1-8}$ a ninth one, $\lambda_9= \sqrt{2/3}\lambda_0$, so that instead of the
$1^-$ octet one  considers a nonet, represented by a non-traceless tensor $G_{\mu\nu}\sim
q_\mu\bar q_\nu$ ($q_i=u,d,s$). Then, to reconcile the physical masses
of $\omega$ and $\phi$ mesons one introduces a mixing angle
$\theta$ and forms combinations like
$\cos\theta \ \omega_0\pm \sin\theta \ \omega_8$  to reproduce the
known masses ($\omega_{0,8}$ are the pure SU(3) $\omega$ meson states).
Alternatively, one can determine the mixing angle from the demand
to reproduce the  quark  content of $\omega$ ($u\bar u + d\bar d$)
and $\phi$ ($s\bar s$) mesons.
The angle $\theta_0$ determined from this condition is
called the ideal mixing angle which is slightly different from $\theta$ obtained from mass formulae. Such a difference
means that in principle the $\phi$ meson can contain a small portion of
non-strange quarks and, vice versa, the wave function of $\omega$ can
contain some hidden strange components. In spite of the fact that the nonet classification does not
have a strict symmetry nature, it  has been found  to excellently describe
the light vector mesons.
However, such a "nonet hypothesis"
needs to be complemented by the restriction that in expressions for
physically observed  processes the trace $G_{\mu\mu}$ will never arise
(see for details, e.g.,  ref.~\cite{okubo77}).

At the level of quark diagrams this restriction means that  topological diagrams with disjoint
parts of quark lines ("hairpin" diagrams) must be zero (in the ideal case) or highly suppressed.
This is known as the Quark-Line Rule or OZI rule. In particular, the quantities
\be
Z_{OZI}=
\frac{\sqrt{2} T(A+B\to X + (s\bar s))} {T(A+B\to X + (u\bar u)) + T(A+B\to X + (d\bar d))}
\label{z} \ee and
\be \beta_{OZI}^2\equiv \left | \frac{\sqrt{2} T(A+B \to X + \phi)} {T(A+B\to X
+ \omega)} \right |^2 = \left |\frac{Z_{OZI}+\tan(\theta-\theta_0)}{1-Z_{OZI}
\tan(\theta-\theta_0)} \right |^2 \label{zz} 
\ee
($A,B$ and $X$ denote non-strange particles, $T$ is the amplitude of the
corresponding process) are  predicted to be small, 
\be |Z|\ll 1, \quad\quad \beta_{OZI}^2\ll 1. \label{ZOZI} 
\ee 
The prediction (\ref{ZOZI}) has been checked for a large
number of experimental cross sections  and is found to be fulfilled with high accuracy
\cite{okubo77}, 
\be 
\beta_{OZI}^2 \le 2.6 \cdot 10^{-2}. \label{beta} 
\ee 
Nevertheless, the OZI
rule being in fact of a mnemonic nature cannot be exact and should have some range of
applicability. So, the unitary condition for the $S$ matrix, $SS^+=1$, implies that 
\be 
2 Im T_{i\to f} \simeq  \sum\limits_{X} T_{i\to X}\ T^*_{f\to  X}, \label{unitary} 
\ee 
where the summation over intermediate states $X$ runs over all open physical channels 
allowed by energy conservation. From (\ref{unitary}) one concludes that at high enough energies
OZI allowed subprocesses $i\to X$ and $f\to X $ always exist
such that they can contribute  to the OZI forbidden $i\to f$ reactions. These correspond to
so-called loop or double hairpin diagrams being topologically equivalent with the forbidden ones.
This means that eq.~(\ref{unitary}) leaves room for OZI rule violation. The experimentally
observed small values of $\beta_{OZI}$ in eq.~(\ref{beta}) may be understood as a random
cancellation of intermediate phases  in (\ref{unitary}) at high energies, so that the OZI rule
can be still fulfilled at high energies \footnote{For elastic processes $i\to i$ the phase
cancellation cannot occur, at any energy, since in this case the
sum in (\ref{unitary}) becomes coherent.}, while at
intermediate energies its violation is always expected.
Nevertheless, at low energies, near the threshold,
the loop diagrams contributing in (\ref{unitary}) can be suppressed
by the lacking energy, and the hairpin diagrams again govern the amplitude of the process.
Hence, since in this kinematical region
there are no other "legal" sources of the OZI rule violation,
an investigation of processes near the threshold is of interest.

Any significant deviation of $\beta_{OZI}$ from (\ref{beta}) near the threshold
indicates  some "exotics" (as the mentioned hidden degrees of freedom)
in the wave functions of the involved particles.
Of particular interest is the presently studied  $\omega$ and $\phi$ production
in nucleon-nucleon reactions since the violation of the OZI rule could
drastically change the interpretation of the quark content of nucleons. 
Nowadays, for the relevant
coupling constants one has the following predictions (cf.\ ref.~\cite{gelmann})
\be
g_{\phi NN}= -\tan (\theta-\theta_0)\ g_{\omega NN}, \nonumber\\
&&
\frac{g_{\rho\pi\phi}^2}{g_{\rho\pi\omega}^2} \simeq (0.7 - 1.) \cdot 10^{-2}, \label{rat}
\ee
where $\Delta\theta = \theta-\theta_0 \simeq  3.7^o \cdots 5.5^o$. Consequently,
within a simplified treatment of the reaction mechanism,
the ratio of the corresponding cross sections
is expected to be proportional (with the proportionality coefficients corrected
by corresponding phase space volumes)
to the ratios of the coupling constants in  eq.~(\ref{rat}), so that possible violations
of the OZI rule are often associated with these values \cite{DISTO1,DISTO2}.
However, the reaction mechanism of $NN\to NNV$, where $N$ denotes the nucleon,
is much more involved and consists of different types of diagrams with quite complicate
interference effects. This hinders a direct investigation
of the validity of the OZI rule; some enhancements  of the ratio (\ref{beta})
may occur dynamically, i.e. the actual ratios of the cross sections may
differ from the "OZI correct" input  ratios of the coupling constants.
Moreover, in processes of the type $pp\to ppV$ the effects of Final State Interaction
(FSI) may become predominant near the threshold and completely mask
the studied problem.

For a reliable study of these effects one needs more experimental
data and more types of processes.  In particular, for further checks of the reaction mechanism
and for a firm separation of FSI effects
it is necessary to  study meson production also at neutron targets.
Near the threshold, FSI in  $pp$ and $pn$ systems
differs due to the Pauli principle, hence a combined study of $pp$ and $pn$
reactions will enlighten the theoretical methods to treat the FSI.
Unfortunately, data on elementary reactions on neutrons are
scarce since they must
be extracted, with some efforts and even mostly with some model
dependent assumptions, from reactions on nuclei, mainly on the deuteron.
The spectator technique \cite{johansson,anke} represents one example of how one can use
a deuteron target to isolate reactions on the neutron.
It is based on the idea to measure the spectator proton, $p_{\rm sp}$, at
fixed beam energy in the vector meson  production reactions
$p d \to d V p_{\rm sp}$, thus exploiting the internal momentum spread of
the neutron inside the deuteron. In such a way one gets
access to quasi-free reactions $p n \to d V$.

An experimental investigation of the near-threshold (pseudo)scalar and vector meson production 
at the neutron becomes therefore feasible. Indeed, at COSY the ANKE spectrometer set-up 
can be used
in particular for studying the $a_0$, $\omega$ and $\phi$ production with the internal beam at
''neutron target'' \cite{anke}. This offers the possibility to enlarge the data base on hadronic
reactions and to address special issues, e.g., for a systematic study of the OZI rule violation
via $\omega$ and $\phi$ production in $\pi N$ and $p p$ reactions (cf.~\cite{sibirtsev} for a
reanalysis) and in $\bar p p$ annihilations 
(cf.~\cite{ellis,rotz} and further references quoted therein for theoretical analyzes) as well. 

OZI rule violations are of interest with respect to possible hints to a significant 
$s \bar s$ admixture in the proton, 
as supported by the pion-nucleon sigma term \cite{donoghue,gasser}
and interpretations of the lepton deep-inelastic scattering \cite{ashman}. Besides the impact on
hadron phenomenology the origin of the OZI rule addresses also a link to QCD \cite{isgur,shuryak}.
Furthermore, the effective description of particle production in elementary processes is a
necessary prerequisite to analyze heavy-ion collisions in detail and to pin  down in-medium
effects. In particular, the $pn$ channels deserve a reliable description which is not simply
accessible from constant isospin factors correcting the cross sections in $pp$ channels.

Given this motivation, in \cite{golubeva,grishina,nakayama} the reaction $p n \to d V$ with $V =
\omega, \phi$ has been studied in some detail ($p n \to d S$ with $S = a_0^+, \eta, \eta'$ is
considered in \cite{golubeva,grishina}). In \cite{golubeva,grishina} the cross sections and
angular distributions are elaborated as a function of the excess energy within a  two-step model.
The same observables are evaluated in \cite{nakayama,nakayama1,ourPhi} within the framework of a
boson exchange model with emphasis on the ratio of cross sections 
$\sigma_{pn \to d \phi} / \sigma_{pn \to d \omega}$ 
being of direct relevance for the OZI rule violation.
Ref.~\cite{nakayama2} focuses on the $\omega$ and $\phi$ meson production in $pp$ reactions. 
We go beyond \cite{nakayama2} by studying also $pn$ reactions and 
by including the deuteron final state.

In this paper we present a combined analysis  of
$\omega$ and $\phi$ meson production in $pp\to ppV$, $pn \to pnV$
and $pn\to dV$ processes. Within an effective meson-nucleon theory we
compute  the covariant amplitudes for $NN\to NNV$, and from
$pp\to ppV$ data we fix the free parameters as to obtain a good description of the
available experimental data. Note that in spite of the large number of parameters
entering the effective meson-nucleon theory
a bulk of them is already determined  by other independent considerations
(the one-boson-exchange (OBE) potential, decays into mesons etc.) 
so that we are left with a restricted number of free parameters
which can be varied. The main idea of the present work is to study
whether it is possible to describe in a consistent way
the $pp\to ppV$, $pn\to pnV$ and $pn\to dV$ amplitudes by exploiting as
input into the calculations such effective parameters
which, at the elementary level, are in a
concord with other data (e.g., known meson and nucleon decays)
and preserve the OSI rule. Then from calculations
of the corresponding cross sections we study  the  possible enhancement of the
respective  ratios and compare with the expected  naive OZI rule predictions.
For a consistent treatment of all mentioned reactions and in order to be able to use
directly the covariant amplitudes from
$NN \to NNV$ processes, with the effective parameters already  found, we perform
the analysis of the $pn\to dV$ processes
within  the Bethe-Salpeter (BS)  formalism \cite{Tjon}  with
the numerical solution \cite{solution} of
the BS equation obtained within the same effective   meson-nucleon theory.
The use of the BS formalism is not dictated by the necessity of taking into
account relativistic effects, rather it is inspired by convenience reasons.

This paper is organized as follow.
In Section \ref{secOdin} the vector meson production in nucleon-nucleon interaction is analyzed.
The kinematics, notation and explicit expressions for the  relevant quantities are presented in
Section \ref{subsecOdin},
the choice of the effective parameters is discussed in Section \ref{subsecDva},
and in Section \ref{subsecTri} results of numerical calculations
of angular distributions and the energy dependence of the  total cross sections for $\omega$ and
$\phi$ meson production in $pp$ and $pn$ reactions are presented. Basing on  the obtained results
the ratio of cross sections $\phi/\omega$ is analyzed in connection with the OZI rule. A similar
structure has the Section \ref{Deu}, where the meson production in the $pn\to dV$ process is
analyzed within the Bethe-Salpeter formalism.   In Section \ref{subDeu1} details of derivation of
the corresponding formulae within the BS formalism are presented. In Section~\ref{subDeu2} results
of numerical calculations of the angular distributions, total cross sections and OZI rule are
discussed. Conclusions and the summary are collected in Section \ref{sumary}.

\section{ The processes $\mbox{\boldmath$ NN \to NN \omega$}$
and $\mbox{\boldmath$NN \to NN \phi$}$} \label{secOdin}

\subsection{Kinematics and Notation}
\label{subsecOdin}

Consider  the vector meson production in $NN$ collisions of the type
\be
N_1+N_2 \to N_1'+N_2'+V. \label{reac1}
\ee
The invariant five-fold cross section is
\begin{eqnarray}
&&
d^5\sigma = \frac{1}{2\sqrt{s(s-4m^2)}} \frac14 \sum\limits_{s_1,s_2}\
\sum\limits_{s_1',s_2',\CM_V}
\,|T_{s_1s_2,s_1',s_2'}^{\CM_V } |^2 d^5\tau_f \ \frac{1}{n! }, \label{crossnn}
\end{eqnarray}
where $s_i$ and $\CM_V$ are the projections of the
nucleon and meson spins on the quantization axis, and
the factor $\displaystyle\frac{1}{n!}$ accounts for $n$ identical
particles in the final state. The invariant phase space volume $d\tau_f$ is defined as
\begin{eqnarray}
d^5\tau = (2\pi)^4\delta\left(p_1+p_2-p_1'-p_2'-q\right)\frac{d^3p_1'}{2E_{{\bf p}_1'}(2\pi)^3}
\frac{d^3p_2'}{2E_{{\bf p}_2'}(2\pi)^3}\frac{d^3q}{2E_{{\bf q}}(2\pi)^3}. \label{tau}
\end{eqnarray}
In eqs.~(\ref{crossnn}) and (\ref{tau}) the 4-momenta of initial  ($p_1, p_2$) 
and final ($p_1', p_2'$) nucleons and vector meson ($q$) are 
$p=(E_{\bf p},{\bf p})$ with $E_{\bf p}=\sqrt{m^2+{\bf p}^2}$,
$q=(E_{\bf q},{\bf q})$ with $E_{\bf q}=\sqrt{m_V^2+{\bf q}^2}$,
where
$m$ and $m_V$ are the nucleon and meson masses, respectively. The initial energy squared of
incident nucleons is defined as  $s=(p_1+p_2)^2$. It is seen from (\ref{tau}) that the cross
section eq.~(\ref{crossnn}) is determined by five independent kinematical variables, the actual
choice of which depends upon the goals of the attacked problem. In the present paper we are
interested in studying the angular distributions of the produced mesons in the center of mass
(CM) of initial particles and the energy dependence of the total cross section. For this sake
it is convenient to choose the kinematics with two invariants, $t=(p_1-q)^2$ and
$s_{12}=(p_1'+p_2')^2$, and two angles, the solid angle $d\Omega_{12}^*$ in the CM of the two
final nucleons and the azimuthal angle  $\varphi_V$ of the meson in the CM of initial particles.
(This is the Chew-Low kinematics.) Then
\begin{eqnarray}\raggedright
d^5\sigma=
\frac{1}{64s(2\pi)^5\sqrt{s(s-4m^2)}}
\sqrt{ \frac{1-{4m^2}/{s_{12}}}{1-{4m^2}/{s}}}
\frac14 \sum\limits_{spins}\
\,|T_{s_1s_2,s_1',s_2'}^{\CM_V } |^2 dt d s_{12}d\varphi_V d\Omega_{12}^* \ \frac{1}{n! }.
\end{eqnarray}
The invariant amplitude $T_{s_1s_2,s_1',s_2'}^{\CM_V }$ is evaluated
within a meson-nucleon  theory based on
effective interaction Lagrangians  which includes
scalar ($\sigma$), pseudoscalar ($\pi$), and neutral ($\omega, \phi$) and charged ($\rho$)
vector mesons (see e.g. \cite{bonncd,chung})
\begin{eqnarray}
{\cal L}_{\sigma NN}&=& g_\sigma \bar N  N \it\Phi_\sigma, \label{eq.9.1}\\
{\cal L}_{\pi NN}&=&
-\frac{f_{\pi NN}}{m_\pi}\bar N\gamma_5\gamma^\mu \partial_\mu
({\boldtau \boldPhi_\pi})N,\\
{\cal L}_{\rho NN}&=& -g_{ \rho NN}\left(\bar N \gamma_\mu{\boldtau}N{\boldPhi_
\rho}^\mu-\frac{\kappa_\rho}{2m} \bar
N\sigma_{\mu\nu}{\boldtau}N\partial^\nu{\boldPhi_\rho}^\mu\right),\\
{\cal L}_{V NN}&=&
-g_{V  NN}\left(
\bar N \gamma_\mu N {\it\Phi}_{V}^\mu-
\frac{\kappa_{V}}{2m}
\bar N \sigma_{\mu\nu}  N \partial^\nu \it\Phi_{V}^\mu\right);
\label{mnn}\\
{\cal L}_{\rho\pi V}&=& g_{\rho\pi V}\,\varepsilon_{\mu\nu\alpha\beta}
\,\partial^\mu\it\Phi_{V}^\nu \,{\rm Tr}\left(\partial^\alpha\it\Phi_\rho^\beta
\it\Phi_\pi\right), \label{eq9}
\end{eqnarray}
where $N$ and $\it\Phi$ denote  the nucleonic and mesonic fields, respectively,
${\rm Tr}\left(\it\Phi_\rho\it\Phi_\pi\right)=\it\Phi_\rho^0\it\Phi_\pi^0
+ \it\Phi_\rho^+\it\Phi_\pi^- + \it\Phi_\rho^-\it\Phi_\pi^+$ and
bold face letters stand for isovectors.
All coupling constants with off-mass shell
particles are dressed by monopole form factors
$F_m=\left(\Lambda^2_m-M^2_m\right)/\left(\Lambda^2_m-k^2_m\right)$,
where $k^2_m$ is the 4-momentum of the virtual particle
with mass $M_m$ ( $k^2_m\neq M_m^2$).

These Lagrangians result into two types of Feynman diagrams: 
(i) the ones which describe the meson
production from the processes of one-boson exchange (OBE) between two nucleons accompanied by the
emission of a vector meson from nucleon lines in $VNN$ vertices (in what follows we call these
diagrams bremsstrahlung type reactions), and 
(ii) production of vector mesons resulting from a
conversion of virtual (exchange) $\pi $ and $\rho$ mesons into a real vector meson, 
i.e., from the $\rho\pi V$ vertex defined by eq.~(\ref{eq9})) 
(these diagrams are called internal conversion type diagrams). 

It is convenient to define for the considered processes an  effective scattering
operator $\hat O(12;1'2'V)$ which can be derived from the Feynman diagrams for the amplitude
by cutting the external nucleon spinors,
as depicted in fig.~\ref{pict1}, where the external nucleons are represented
by solid truncated lines. All the dependencies on meson and nucleon propagators as well as
on the polarization of the final meson are included into the definition of $\hat O(12;1'2'V)$.
Accordingly, the invariant amplitude
$T_{s_1s_2,s_1',s_2'}^{\CM_V }$ can be obtained by merely sandwiching
this operator between the nucleon spinors. For example, for the $pp$ reaction one has
\begin{eqnarray}
T_{s_1s_2,s_1',s_2'}^{\CM_V }=
\left \langle p_1',s_1', p_2',s_2'\ \left | \ \hat O(12;1'2'V \ \right |\  p_1,s_1, p_2,s_2\ \right\rangle
-{\rm antisym.} \label{ampl}
\end{eqnarray}
In the case of $pn$ processes, instead of the antisymmetrization in eq.~(\ref{ampl}), denoted by
"antisym", one should properly take into account the isospin factors (see for details
ref.~\cite{titov}).
The derivation of the analytical form of the corresponding
bremsstrahlung and conversion amplitudes (\ref{ampl})
from the Feynman graphs in  fig.~\ref{pict1} from the
Lagrangians (\ref{mnn}) is
straightforward but rather cumbersome and we do not present it here.
Explicitly the amplitudes $T_{s_1s_2,s_1',s_2'}^{\CM_V}$
can be found elsewhere, e.g., in ref.~\cite{titov}.

\subsection{Effective Parameters}\label{subsecDva}

\begin{figure}[h]  %      Fig1
\includegraphics[width=0.999\textwidth]{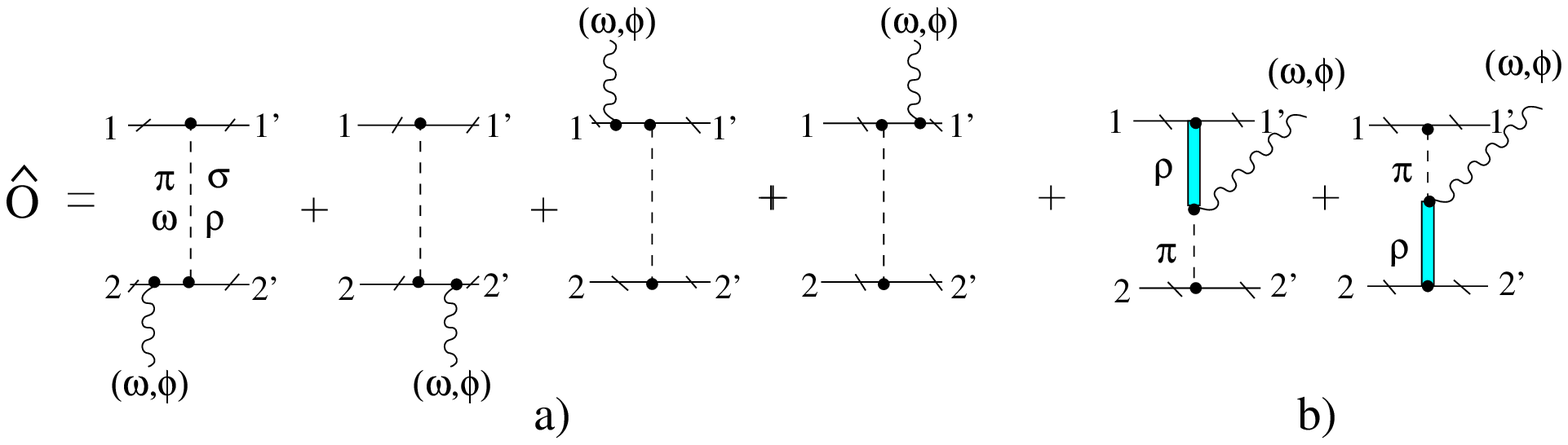} %
\caption{ Graphical representation of the  scattering operator $\hat O(12;1'2'V)$ (with truncated
nucleon lines) for the vector meson production in $NN$ interactions within an
effective meson-nucleon theory. The dashed and solid thick lines symbolize the OBE of
$\pi,\sigma, \omega, \rho$ mesons with effective parameters listed in Table~\ref{tb1} 
(see also ref.~\cite{bonncd}), 
the wavy lines represent the produced vector meson. The first group of
diagrams (a) corresponds to meson production from bremsstrahlung processes, while the last two
diagrams (b) depict meson production from the internal
conversion of the virtual $\pi \rho$ into real $\omega$ or $\phi$ mesons.
The dots symbolize the dressing of the corresponding
lines of virtual particles with monopole cut-off form factors. }
\label{pict1}
\end{figure}

In this subsection we discuss the choice of effective parameters for the
bremsstrahlung and conversion processes. The bremsstrahlung diagrams
(fig.~\ref{pict1}a)
consist of two parts, the pure OBE exchange and the emission  of the meson
from the nucleon lines. The OBE parameters, masses, cut-off's and coupling constants,
are adopted to coincide with those of the OBE potential by  the
Bonn group with the pseudo-vector coupling in the $\pi NN$ vertex~\cite{bonncd}.
Then in the bremsstrahlung part of the amplitude (\ref{ampl}) we are left
with the parameters of the $VNN$ vertices (vertices with wavy lines in fig.~\ref{pict1}a)
for the real production of $\omega$ and $\phi$. In principle,
the effective parameters for such vertices can differ from the corresponding parametrization
of the virtual mesons; consequently in the present work they are considered as free parameters to
be fitted independently. For the conversion type vertices $\rho\pi V$ (vertices with wavy lines in
fig.~\ref{pict1}b) the coupling constant $g_{\rho\pi\omega}$ has been fixed from a systematic
study of pseudoscalar and vector meson radiative decays together with the  vector meson dominance
conjecture (see, e.g. \cite{nakayama,durso,chung})  and is found to be $g_{\rho\pi\omega}\simeq
12.9\ GeV^{-1}$. ( At this point it is worth mentioning that often in the literature the
definition of the coupling constants $g_{\rho\pi V}$
differs from ours in  eq. (\ref{eq9})  by a mass  factor  included
into $g_{\rho\pi V}$ as to make it dimensionless. So, in refs.~\cite{nakayama,dashen} this
factor reads as $ \sqrt{m_\rho m_\omega}$ which results in  a dimensionless coupling constant
$g_{\rho\pi\omega}\simeq 10.0$.) The corresponding
cut-off parameter, also freely adjustable in the present consideration, has been
adjusted to reproduce the angular distribution of $\omega$ mesons 
from COSY-TOF experiments \cite{TOF} yielding 
$\Lambda_{\rho\pi\omega} \simeq 1000 \ MeV$.
For the $\phi$ production from conversion type diagrams the coupling constant $g_{\rho\pi\phi}$
is adjusted to data for the $\phi \to \rho\pi$ decay. The value
$\Gamma(\phi\to\rho\pi)=0.69 \ MeV$ suggests $g_{\rho\pi\phi} = -1.1 \ GeV^{-1}$ \cite{titov} 
(cf.\ also ref.~\cite{chung}). 
Then for the conversion diagram the cut-off parameter $\Lambda_{\rho\pi\phi}$
is found from a combined analysis of the contributions of the conversion and bremsstrahlung
diagrams to the experimental cross section \cite{DISTO1,DISTO2}. Together with the nucleonic
cut-off $\Lambda_N = 720\ MeV $ we
adopt $\Lambda_{\rho\pi\phi} = 1340 \ MeV$, following ref.~\cite{titov} (set "B").

Next we explain the choice of parameters for the bremsstrahlung part. Since the nucleon cut-off
parameter, $\Lambda_N$, has already been fixed, we have now to choose only the coupling
constant $g_{\omega NN}$ (and $\kappa_\omega$) for the emission of the $\omega$ meson. The
corresponding parameters for the $\phi$ meson are connected with symmetry relations with the
$\omega$ meson and we do not consider here them as free parameters. The cut-off for $\omega $
meson  production as well as for the $\phi$ meson
is chosen as in the OBE potential \cite{bonncd}
($\Lambda_{\phi NN}=\Lambda_{\omega NN}$).
In the present work the coupling constant $g_{\omega NN}=9.0$ \cite{nakayama} is chosen. Then,
having fixed the $\omega$ parameters, we find $g_{\phi NN} = -g_{\omega NN} \tan\Delta\theta
\simeq - 0.8$ with $\Delta\theta \approx 5^o\ $ \cite{okubo77} (cf. also ref.~\cite{titov}).
The parameters used in our calculations are listed in Table~\ref{tb1}. It should
be noted that the chosen parameters are in the range quoted by eq.~(\ref{rat}).

\begin{table}[!ht]
\caption{Model parameters for the process  $NN\to NNV$ ($V=\omega, \phi$).}
\begin{tabular}{l c l c l }
\hline\hline  % after \\: \hline or \cline{col1-col2} \cline{col3-col4} ...
Vertex                    &\phantom{ppp} &     Coupling constant           &\phantom{ppppppp} &             Cutoff      \\
                          &              &($f_{VNN}=\kappa_{V}g_{VNN}$ )   & &              (MeV)        \\\hline
$\rho\pi V$               & & $g_{\rho\pi\omega}=12.9 \ GeV^{-1}$ \cite{nakayama,moya}     & &       $\Lambda_{\rho\pi\omega} =980$ \cite{nakayama} \\
                          & &$g_{\rho\pi\phi}=-1.1 \ GeV^{-1}$ \cite{titov,nakayama1}& &$\Lambda_{\rho\pi\phi} =1340$ \cite{titov}    \\
                          & &                                     & & $\Lambda_{ NN} =720$                            \\
$V NN$                    & & $g_{\omega NN}=9.$                  & & $\Lambda_{\omega NN}=1500$\\
                          & &  ($\kappa_\omega=0.5$)              & &                           \\
                          & &  $g_{\phi NN}=-0.8$ \cite{titov} & & $\Lambda_{\phi NN} =\Lambda_{\omega NN}=1500$\\
                         & &  ($\kappa_\phi=0.5 $, $\Delta\theta\approx 5^o$ \cite{okubo77})                & &                           \\
$MNN$\footnote{The OBE parameters could, in principle, be different
from the ones in real production of on-mass shell mesons from $NNV$ vertices.} & &                                 & &                           \\
    $\pi $                & & $f_{ \pi NN}=1.0$               & & $\Lambda_{ \pi NN} =1300$    \\
    $\sigma$              & & $g_{\sigma NN}=10.$              & & $\Lambda_{ \sigma NN} =1800$  \\                          \\
    $\rho$               & & $g_{\rho NN}=3.5$                & &$\Lambda_{ \rho NN} =1300$     \\
                         & & ($\kappa_\rho=6.1$)              & &                  \\
    $\omega$             & &  $g_{\omega NN}=15.85$           & &$\Lambda_{ \omega NN} =1500$     \\
                         & & ($\kappa_\omega=0.0$)            & &                           \\
\hline\hline
\end{tabular}
\label{tb1}
\end{table}

\subsection{Results for $\mbox{\boldmath$NN \to NNV$}$} \label{subsecTri}

\begin{figure}[h]  %      Fig2
\raggedright
\begin{minipage}{8cm}
\includegraphics[width=0.9\textwidth]{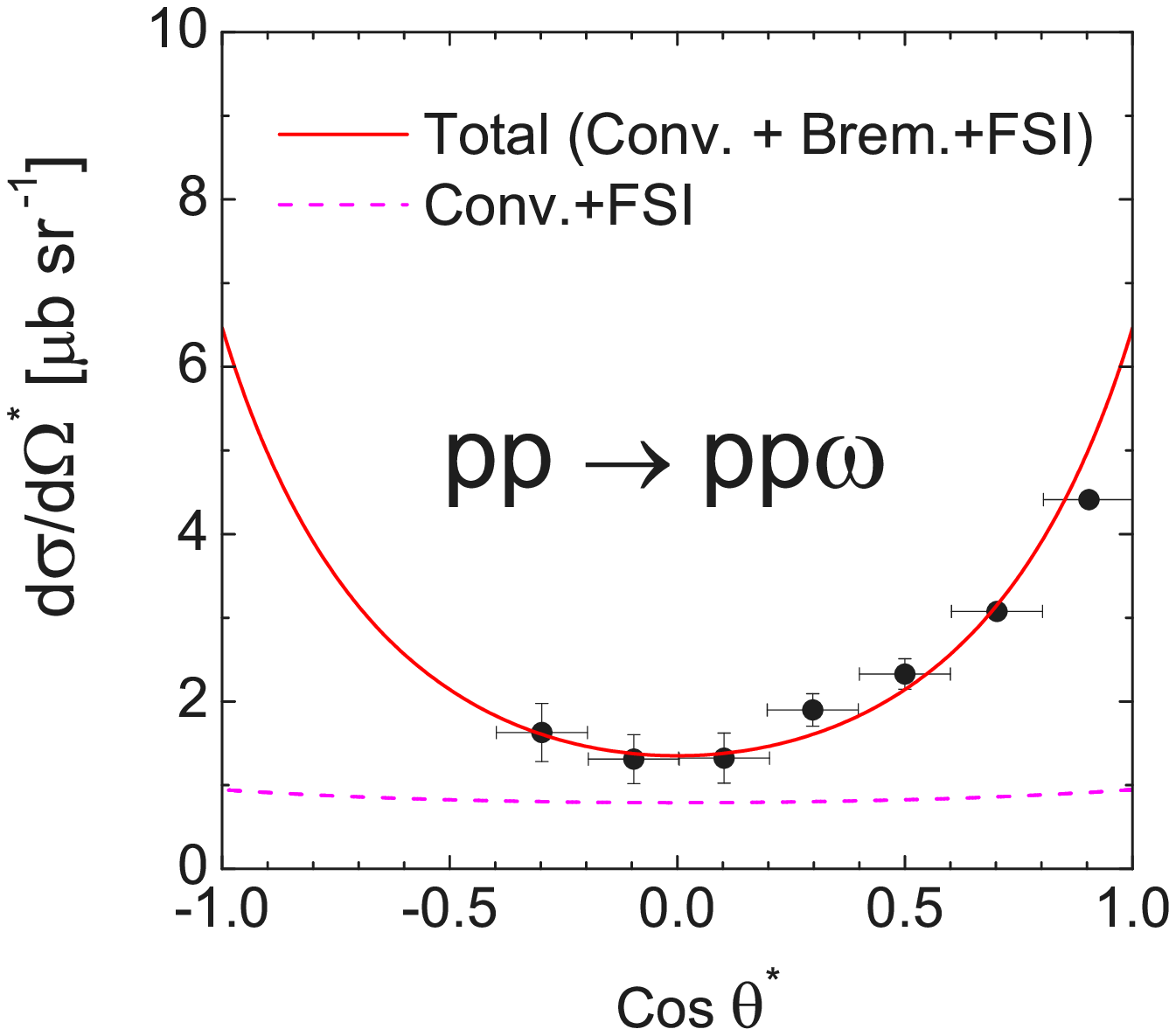} %
 \end{minipage}

\vskip -6.4cm
\raggedleft
\begin{minipage}{8cm}
\includegraphics[width=0.9\textwidth]{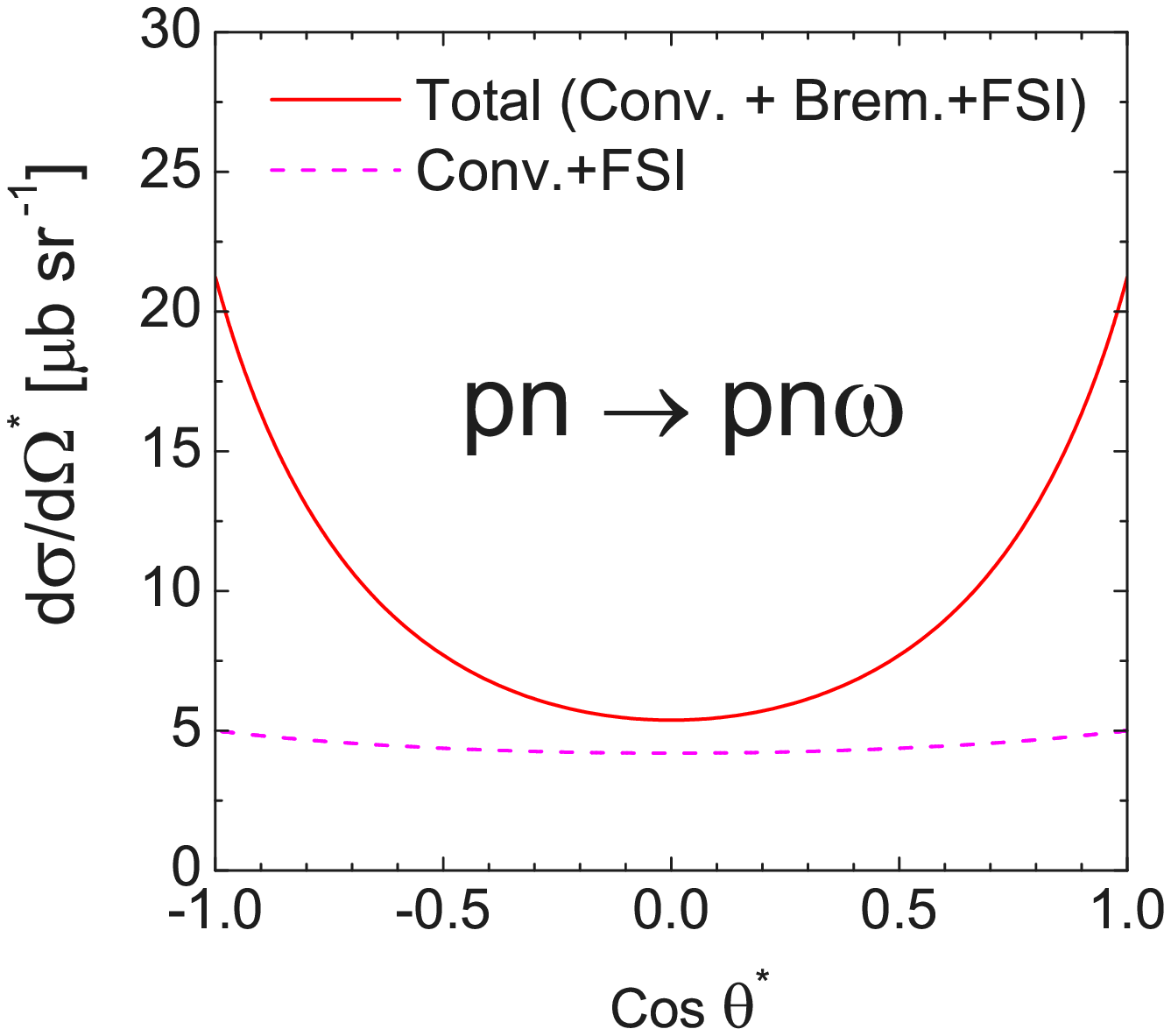}
\end{minipage}
\caption{ Angular distributions of $\omega$ mesons for the reaction $NN\to NN\omega$ at
the excess energy  $\Delta s^{\frac12}=173 \ MeV$. Dashed lines correspond to contributions  of
conversion currents (fig.~\ref{pict1}b), solid lines represent the total contributions of
bremsstrahlung and conversion diagrams (figs.~\ref{pict1}a and \ref{pict1}b). FSI is included in
all contributions. Experimental data are from the COSY-TOF Collaboration \cite{TOF}. }
\label{pict2}
\end{figure}

\begin{figure}[h]  %
\includegraphics[width=0.6\textwidth]{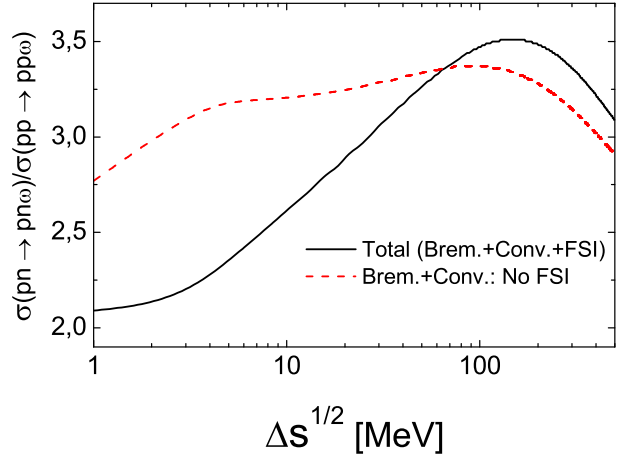} %
\caption{Ratios of the total cross sections of $\omega$ meson production in $pp$ and $pn$
channels as a function of the excess energy.
Dashed line denotes the ratio of cross sections without FSI; solid line reflects the
ratio of total cross sections with FSI taken into account. } \label{pict4}
\end{figure}

In our calculations we make use of the explicit expressions for the conversion and bremsstrahlung
diagrams quoted in ref.~\cite{titov}. The FSI effects have been calculated within the Jost
function formalism \cite{gillespe} which reproduces the singlet and triplet phase shifts at low
energies. In fig.~\ref{pict2} we present results of calculations of the angular distribution of
$\omega$ mesons at the excess energy $\Delta s^{1/2} = 173 \ MeV$ ($\Delta s^{1/2}\equiv
\sqrt{s}-2m_p-m_V$) for proton (left panel) and neutron (right panel) targets. The experimental
data~\cite{ankeExper} served to fix our free parameters for further calculations of the energy
dependence of the total cross section and for an analysis of $pn\to dV$ processes. It can be seen
that with the set of parameters listed  in Table~\ref{tb1}
a good description of the data is achieved. It is also found that the contribution of
bremsstrahlung diagrams (fig.~\ref{pict1}a) is predominant
in both $pp$ and $pn$ processes (see fig.~\ref{pict4}). The difference in magnitudes for
$pp$ and $pn$ processes is due to  the Pauli principle for the
former reactions (integration over
$d\Omega_{12}^*$  in eq. (\ref{crossnn}) is performed only over
one hemisphere of the two protons phase volume) and a possible destructive
interference of diagrams due to antisymmetrization effects and
isospin enhancements for the latter (see for details ref.~\cite{titov}).

\begin{figure}[h]  %      Fig3
\raggedright
\begin{minipage}{8cm}
\includegraphics[width=1.0\textwidth]{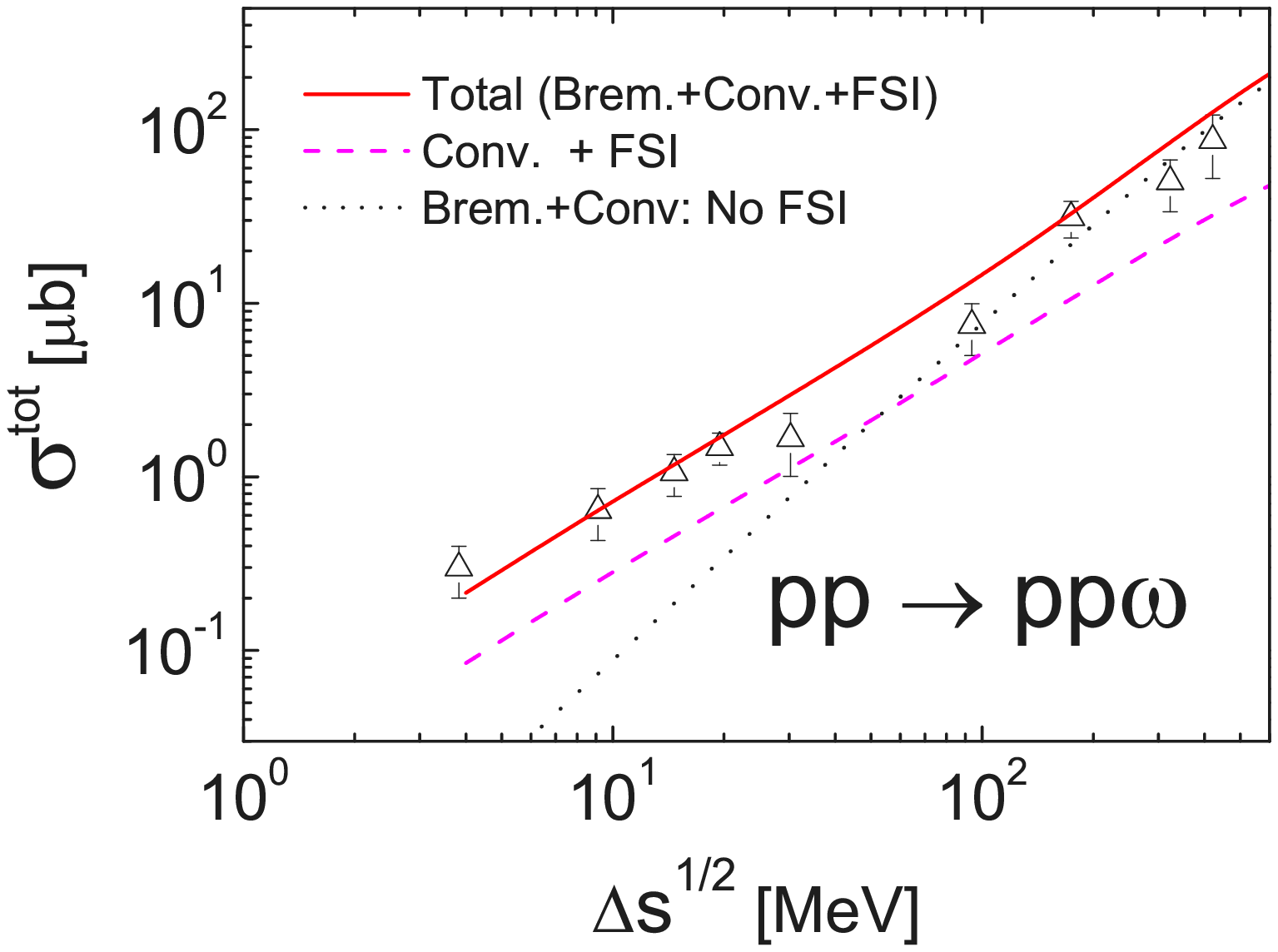} %
\end{minipage}

\vskip -7.0cm \raggedleft
\begin{minipage}{8cm}
\includegraphics[width=1.18\textwidth]{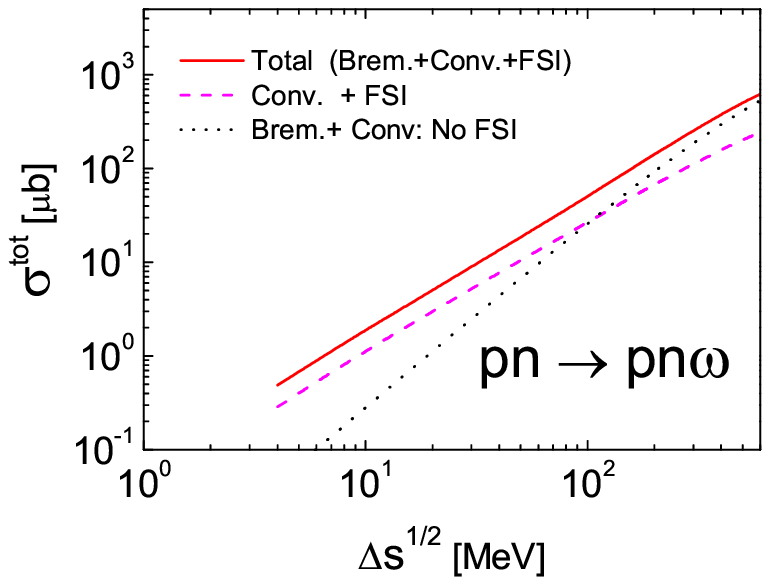}
\end{minipage}
\caption{Total cross sections of near-threshold production of  $\omega$ mesons
in the process  $NN\to NN\omega$ (left panel: $pp$ reaction, right panel: $pn$ reaction)
as a function of the energy
excess $\Delta s^{\frac12}=\sqrt{s} - 2m_p - m_\omega$.
Dashed lines correspond to contributions of conversion currents solely (fig.~\ref{pict1}b),
dotted and solid lines represent  results with including bremsstrahlung and conversion  diagrams
(figs.~\ref{pict1}a and \ref{pict1}b) without and with FSI effects taken into account,
respectively. Experimental data are from SATURNE \cite{Hibou}, COSY-TOF Collaboration \cite{TOF}
and DISTO Collaboration \cite{DISTO1,DISTO2} (cf.\ the compilation in \protect\cite{nakayama2}).}
\label{pict3}
\end{figure}

\begin{figure}[h]  %      Fig5
\raggedright
\begin{minipage}{8cm}
\includegraphics[width=0.9\textwidth]{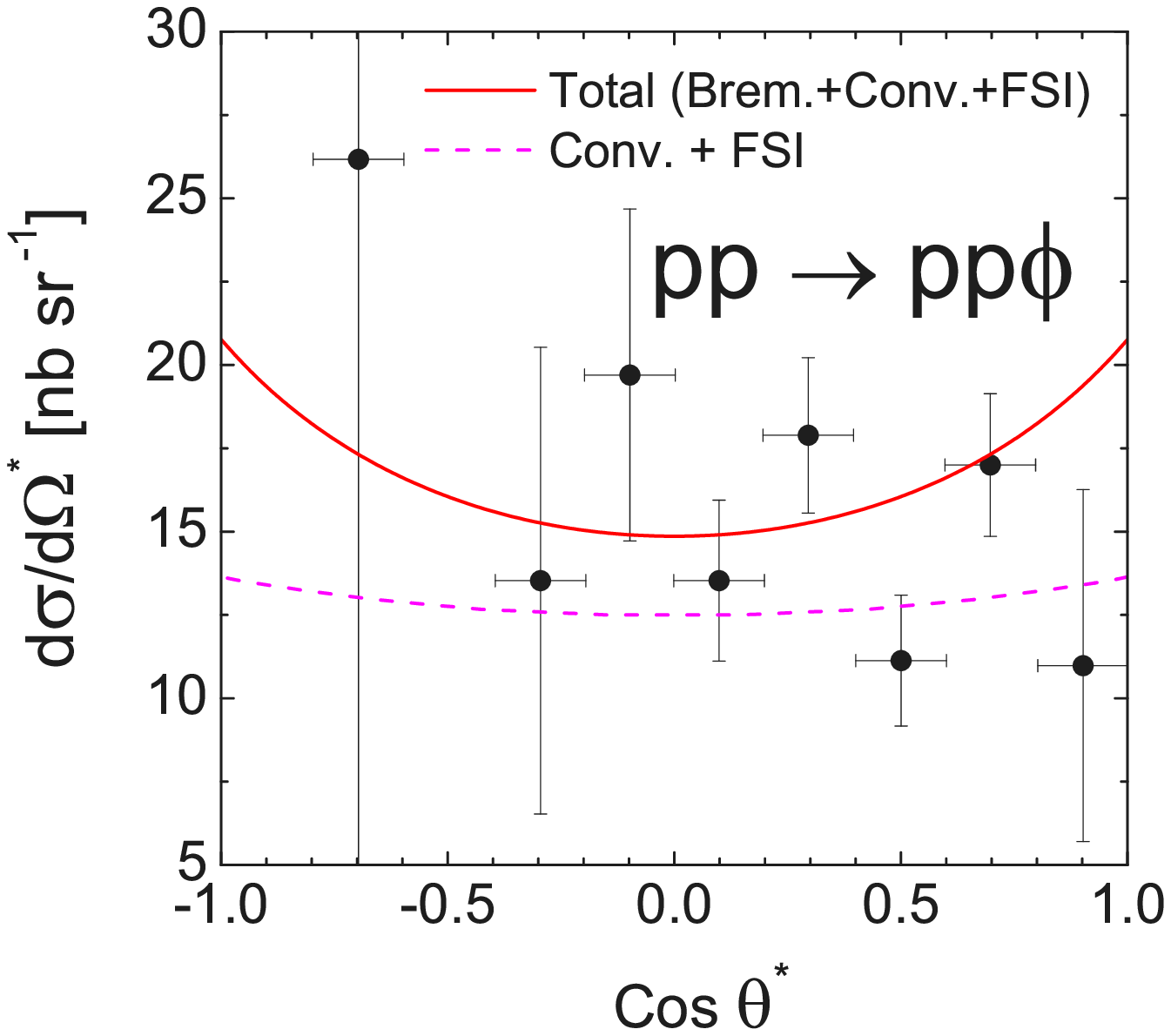} %
\end{minipage}

\vskip -6.4cm
\raggedleft
\begin{minipage}{8cm}
\includegraphics[width=0.9\textwidth]{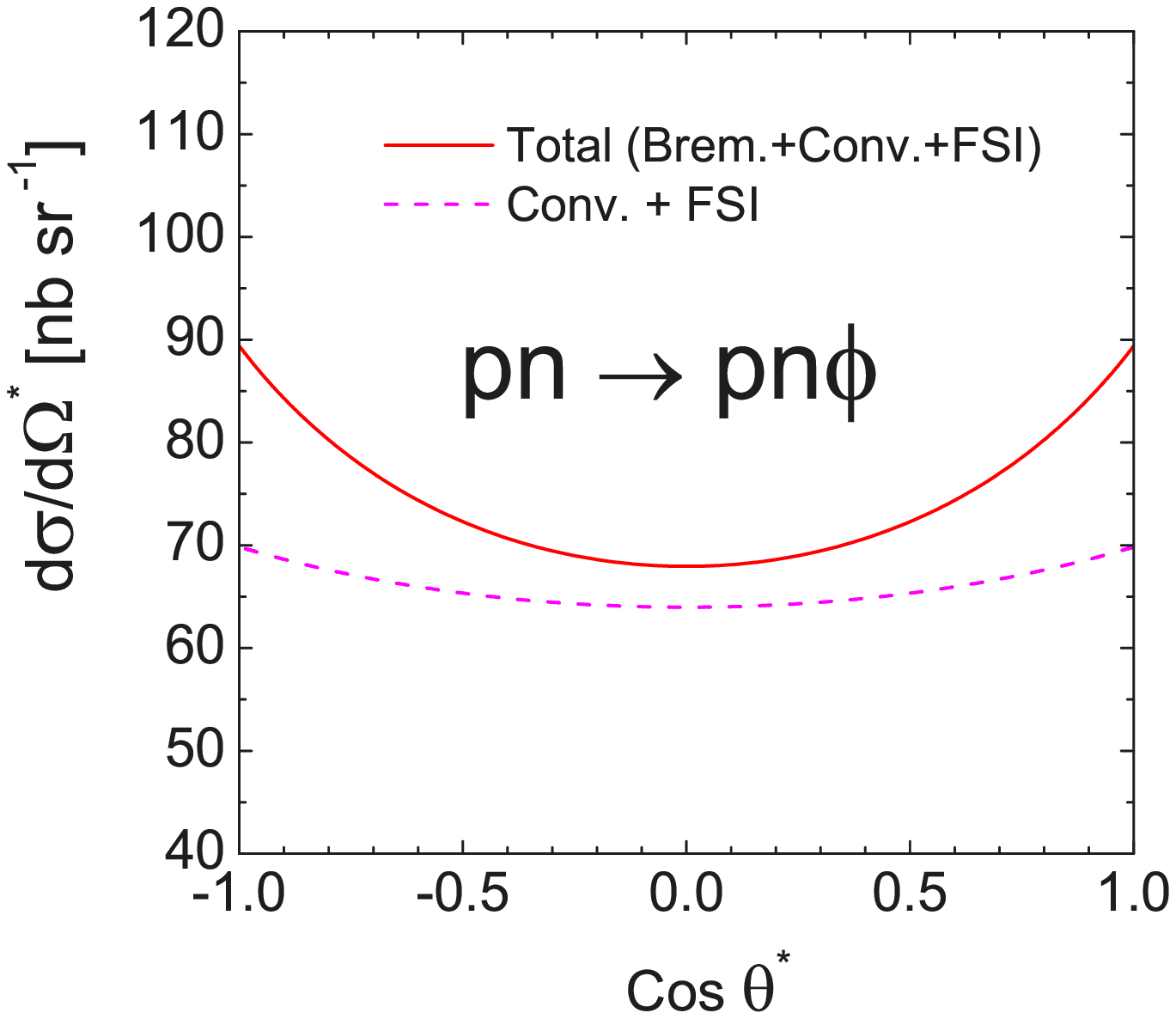}
\end{minipage}
\caption{Angular distributions of $\phi$ mesons in the process $NN\to NN\phi$ (left panel: $pp$
reaction, right panel: $pn$ reaction) at the energy excess $\Delta s^{\frac12}=83 \ MeV$.
Experimental data are from the DISTO Collaboration \cite{DISTO1,DISTO2}. Notation as in
fig.~\ref{pict2}.} \label{pict5}
\end{figure}

\begin{figure}[h]  %      Fig6
\raggedright
\begin{minipage}{8cm}
\includegraphics[width=1.02\textwidth]{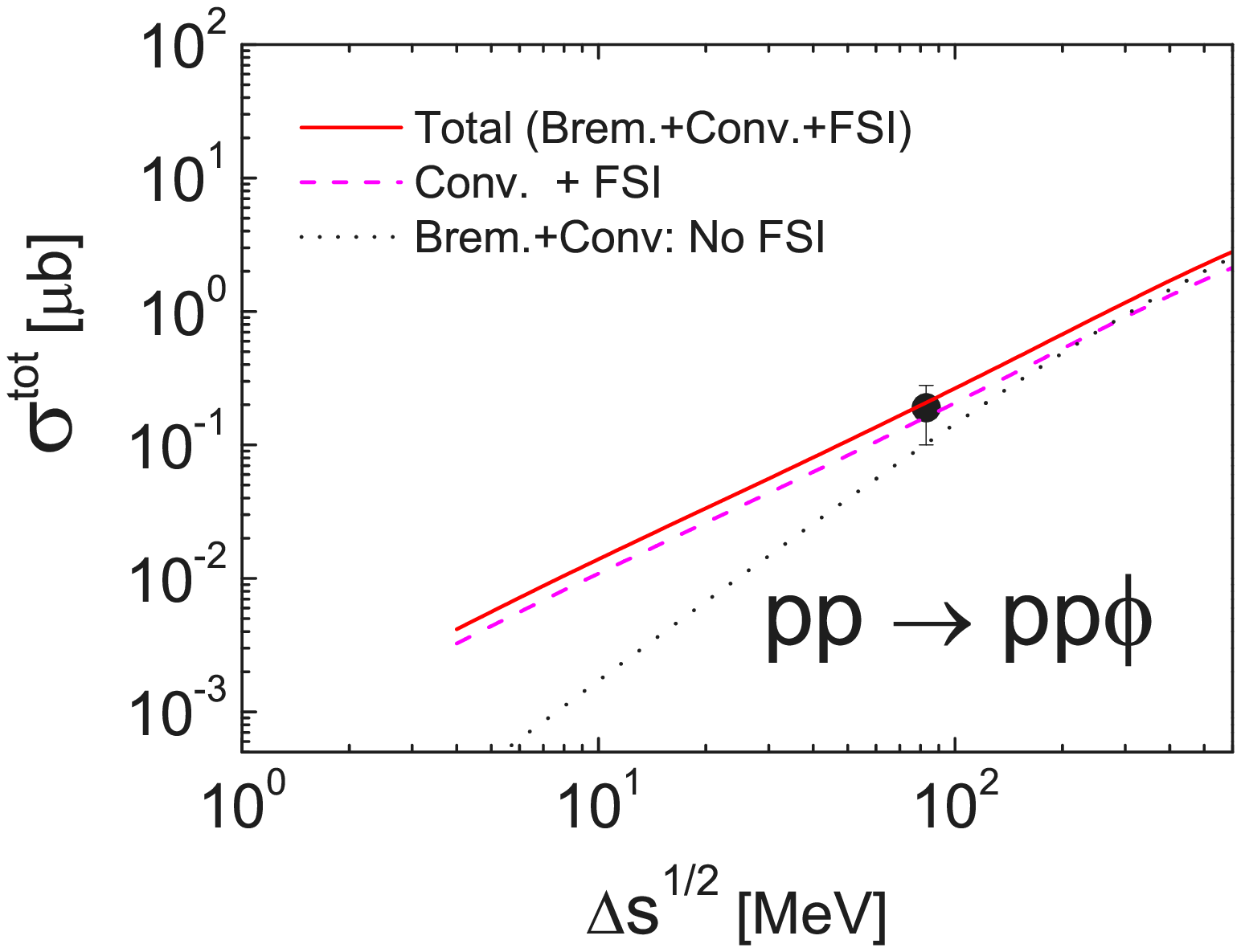} %
\end{minipage}

\vskip -6.2cm
\raggedleft
\begin{minipage}{8cm}
\includegraphics[width=1.02\textwidth]{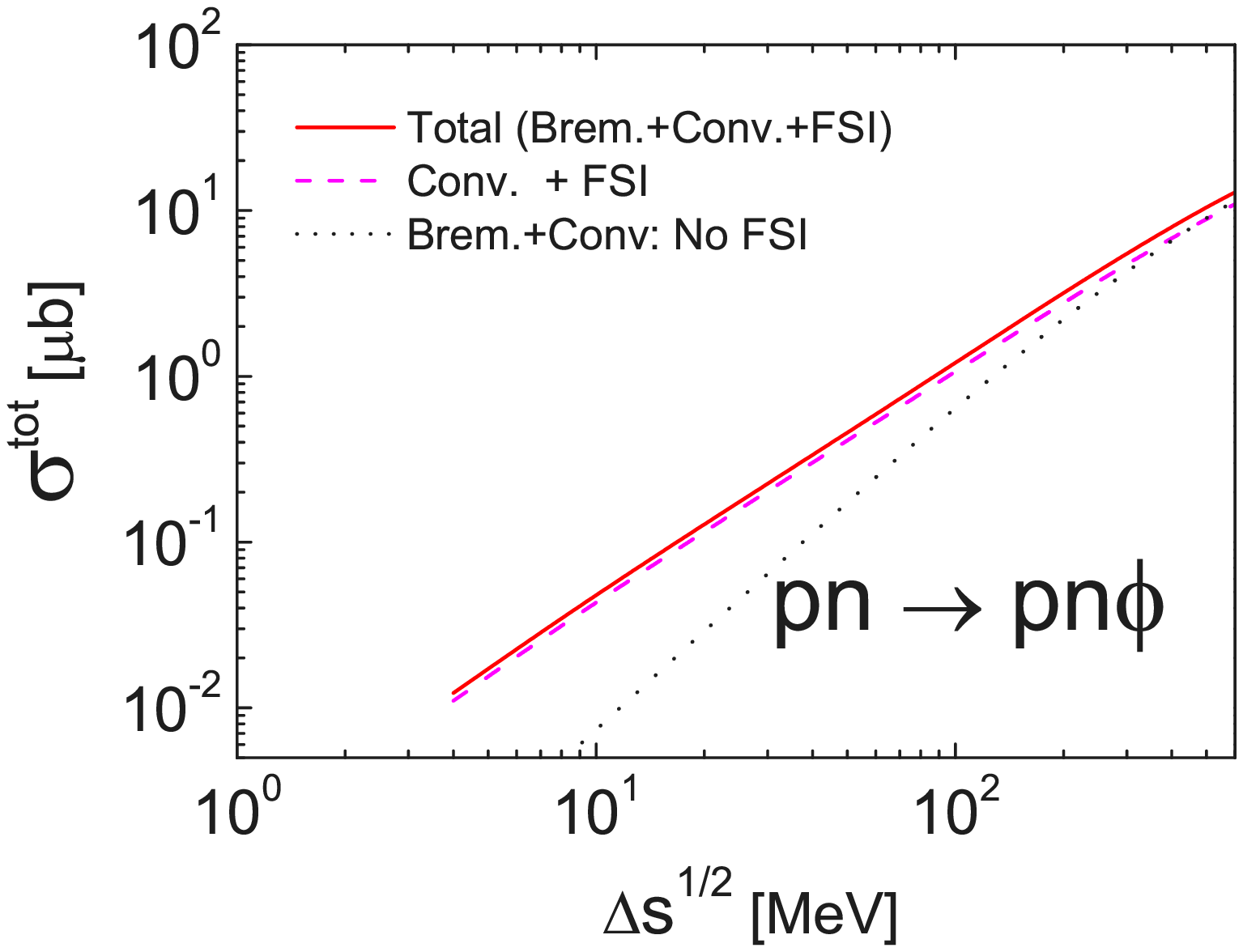}
\end{minipage}
\caption{
Total cross sections of near-threshold production of  $\phi$ mesons
in the process $NN\to NN\phi$ as a function of the energy
excess $\Delta s^{\frac12}=\sqrt{s} - 2m_p - m_\phi$. Experimental data is from
the DISTO Collaboration \cite{DISTO1,DISTO2}. Notation as in fig.~\ref{pict3}.} \label{pict6}
\end{figure}

\begin{figure}[h]  %
\includegraphics[width=0.6\textwidth]{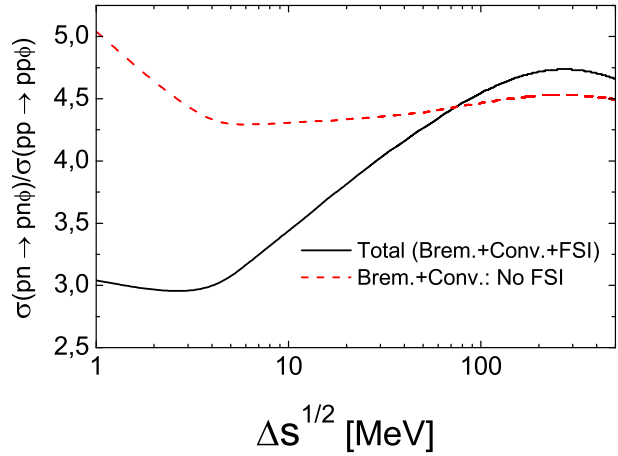} %
\caption{Ratios of the total cross sections of $\phi$ meson production in $pp$ and $pn$ channels.
Notation as in fig.~\ref{pict4}.} \label{pict7}
\end{figure}

The remaining parameters for $\phi$ have been fixed to describe the data for the angular
distribution of $\phi$ production at $\Delta s^{1/2} = 83 \ MeV$ \cite{DISTO1,DISTO2}.
Having adjusted our parameters  in this manner  we compute
the energy dependence of the total cross section. Results of calculations together
with the available experimental data are presented in
figs.~\ref{pict3}, \ref{pict5} and \ref{pict6}. The dashed lines
represent the contribution  of the conversion diagrams alone, the dotted
lines are results of calculations of both bremsstrahlung and conversion diagrams
by neglecting the FSI effects. Eventually, the solid lines exhibit
the total cross section  with FSI taken into account \cite{titov}.
A fairly good description of the  data in a large interval of the excess energy is achieved.
Contrary to $\omega$ production, the contribution of the bremsstrahlung
diagram for $\phi$ mesons is much smaller, due to the small value of the $g_{\phi NN}$ coupling.
The nontrivial behavior of the ratio $\sigma_{pn\to pnV}/\sigma_{pp\to pp V}$ exhibited in
figs.~\ref{pict4} and \ref{pict7} evidence that  the vector meson production cross sections known
by $pp$ reactions can not simply be translated into ones for $pn$ reactions. A comparison of the
solid and dashed lines in figs.~\ref{pict4} and \ref{pict7} clearly indicates
that near the threshold the FSI in $pp$ and $pn$ systems is different.
Rather, a profound analysis, like the one presented here, is needed to have reliable input for
simulating heavy-ion and proton-nucleus collisions. This is particularly important 
for the ongoing experiments with HADES \cite{HADES}.

\begin{figure}[h]  %      Fig8
\includegraphics[width=0.6\textwidth]{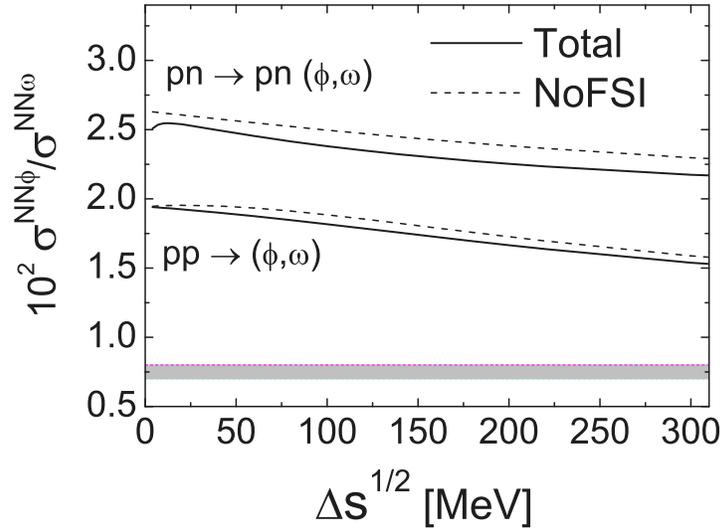} %
\caption{Ratio of the total $\phi$-to-$\omega$ production cross sections in $pp$ and $pn$
reactions. Dashed lines denote the ratio of "net" cross sections, i.e. without any FSI; solid
lines reflect the ratio of total cross sections with FSI taken into account. The shaded area
indicates the range of ratios
$\left( \displaystyle\frac{g_{\rho\pi\phi}}{g_{\rho\pi\omega}}\right)^2\approx 0.0071$ and
$\left(  \displaystyle\frac{g_{\phi NN}}{g_{\omega NN}}\right )^2\approx 0.0079$
(cf.\ Table~\ref{tb1}) used as input for the calculations.}\label{pict8}
\end{figure}

\begin{figure}[h]  %      Fig9
\includegraphics[width=0.95\textwidth]{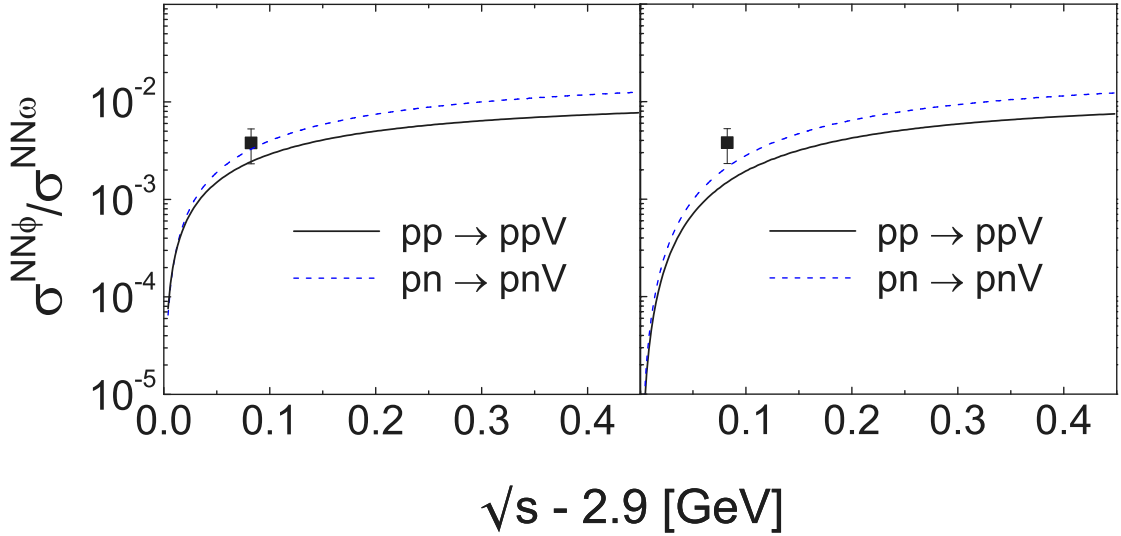} %
\caption{Ratio of the total $\phi$-to-$\omega$ production cross sections in $pp$ (solid lines) and
$pn$ (dashed lines) reactions as a function of the
excess energy above
the $\phi$ threshold. Left (right) panel corresponds to ratio of total cross sections with
(without) FSI effects taken into account. Data for the $pp$ reaction is from DISTO
Collaboration \cite{DISTO2}. } \label{pict9}
\end{figure}

Figures \ref{pict3} and \ref{pict6} demonstrate that
in the extreme threshold limit the main contribution comes from the final state
interaction effects. As the excess energy increases the role of FSI diminishes,
even becoming negligible at excess energy $\Delta s^{1/2} > 200 \ MeV$.
This is a disappointing fact, since as mentioned above, the checks of
the validity of the  OSI rule are preferably to be performed at as
low energies as possible (where the loop and double hairpin diagrams are suppressed),
while we find that at these energies the "net" cross section relevant for the OZI rule
is completely  masked by FSI. However, if
the FSI effects can be firmly separated, the total cross section
may still serve as criterion of the validity of the OSI rule. This could be estimated
by an investigation the OZI  $\phi/\omega$ ratio at
equal excess energies, where the effects of
difference in phase space volumes is minimized,
for  cross sections with and without including FSI corrections.
Figure~\ref{pict8} illustrates that the dependence of such ratios
upon the energy is rather weak and practically
is not affected by the FSI effects, which let us argue that the adopted
Jost method correctly describes the FSI effects.
The small difference between ratios with and without FSI can  be
traced back to the different interactions in $pp$
($^1S_0$ configuration) and $pn$ ($^3S_1 - {}^3D_1$ configurations) systems
near the threshold. The results in fig.~\ref{pict8}  persuade
that in this kinematical region the FSI effects, playing an
important role in the absolute value of the total cross sections,
do not essentially  mask  the study of OZI rule by
ratios of total cross sections. Note that the absolute values
of the ratios depicted in fig.~\ref{pict8}  clearly indicate a
violation of the OZI rule. But this is an apparent effect since
the relative contributions of bremsstrahlung and conversion
diagrams and, correspondingly, the interference
effects are quite different for $\omega$ and $\phi$ production
(cf.\ figs.~\ref{pict3} and \ref{pict6}). Moreover,
even at equal excess energies the phase space volumes for
$\phi$ and $\omega$ mesons are quite different.
Sometimes in the  literature one studies the ratio $\phi/\omega$
at equal initial beam energies \cite{DISTO1}.
In this case one has two effects with opposite contributions:
(i) the phase space volume of the
$\phi$ meson is much smaller in comparison with the $\omega$ meson
(due to different masses, there is a shift of $\sim 240\ MeV$
in $\Delta s^{1/2}$) which will decrease the ratio, and 
(ii) since at equal beam energies the relative momentum of two produced
nucleons is smaller for $\phi$ meson production, the FSI corrections
are expected to be larger (cf.\ figs.~\ref{pict3} and \ref{pict6}).
In fig.~\ref{pict9} the OZI ratio at equal beam energies
is presented together with data from \cite{DISTO1}. It is seen that these ratios
are by an order of magnitude smaller than that at equal excess energy and more
compatible with the naive OZI rule predictions. In the right panel of fig.~\ref{pict9}
the OZI ratio is depicted without any FSI corrections. A comparison with
the left panel and with the range of the ratio of
input constants (the shaded area in
fig.~\ref{pict8}) gives some evidence about the magnitude of
corrections from the phase space volumes solely.

From the performed analysis one can conclude that an investigation
of the OZI rule in $NN\to NNV$ processes near the threshold is feasible,
provided one can firmly take into account the FSI effects
and, consequently, properly calculate the phase space corrections.
The problem of accounting for the FSI between the two nucleons
can be solved by considering processes with a definite two-nucleon
final state, e.g. the process where the FSI in a $pn$ system result
in a bound state, the deuteron.

\section{The processes $\mbox{\boldmath$pn \to d \omega$}$ and
$\mbox{\boldmath$pn\to d \phi$}$}\label{Deu}

One way of testing our assertion on the predicted cross sections for $pn$ reactions is to
implement them in the tagged quasi-free reaction $pn\to dV$.

\subsection{Formalism}\label{subDeu1}

Let us  consider the reaction
\begin{equation}
p\,+n\,=\,d+V,  \label{reaction}
\end{equation}
where, as before, $V$ denotes the $\omega$ or $\phi$ vector meson and $d$ the deuteron.
The invariant differential cross section reads 
\be
\frac{d\sigma}{dt} =
\frac{1}{16\pi\,s(s-4m^2)}\,\frac{1}{4}\sum\limits_{s_1,s_2}
\sum\limits_{\CM_V,\CM_d}
\,|T_{s_1s_2}^{\CM_V\CM_d}(s,t)|^2, \label{eq1} 
\ee 
where $s$ is the square of the total energy of colliding particles,
$t$ is the square of the transferred 4-momentum, $s_1,s_2,\CM_V$ and $\CM_d$ denote
the spin projections on a given quantization axis, and $T$ stands
for the invariant amplitude of the process (\ref{reaction}).
As in the previous section the most general form of the amplitude
$T$ is presented in the form
 \be
T_{s_1s_2}^{\CM_V\CM_d}(s,t)=\langle D,\CM_d\,|\hat
G_\mu\,\xi^{*\mu}_{\CM_V}|1,2\rangle, \label{wq1} 
\ee
where $\xi_{\CM_V}$ is the polarization vector of the vector meson. The scattering
operator $\,\hat G\,\,$ represents a vector in the vector space of mesons and a
$16\otimes 16$ component object in the spinor space of nucleons; the deuteron is
described as a $16$ component BS amplitude $\Phi(1,2)$ which
is defined as a matrix element of a time-ordered product of two nucleon
fields $\psi(x)$ as
\be
\Phi^{\alpha\beta}(1,2)=\langle D| T\left(\psi^\alpha(1)\psi^\beta(2)\right)|0\rangle
\label{eq3}
\ee
and satisfies the BS equation.

Suppose that in the considered reactions the off-mass shell effects are negligibly
small, i.e., the on-mass shell matrix elements of the Lagrangian (\ref{eq9})
between real nucleons (reaction (\ref{reac1})) and  half-off mass shell
matrix elements (reaction (\ref{reaction})) are the same.
Then it can be immediately seen that the operator $G^\mu\,\xi_\mu^*$
coincides with the corresponding operator of the process (\ref{reac1}), i.e.,
$\hat O = G^\mu\,\xi_\mu^*$. The invariant amplitude may be cast in the form
\be
T_{s_1s_2}^{\CM_V\CM_d}(s,t)=-i\int\,\frac{d^4p}{(2\pi)^4}\bar \Phi_{\CM_d}^{\alpha b}(1',2')\,
\hat O^{bc}_{\alpha\beta}(12;1'2',\CM_V)
u^c(1)u^\beta(2),
\label{eq5}
\ee
where $\bar\Phi_{\CM_d}^{ab}(1',2')$ is the conjugate
BS amplitude in the momentum space,
$p$ is the relative 4-momentum of the nucleons in the deuteron, and
$u(1), u(2)$ denote the Dirac spinors for the incident nucleons. Recall that the
operator $\hat O^{bc}_{\alpha\beta}(12;1'2',\CM_V)$ is a scattering operator describing
the vector meson production in the final state. This operator acts in the spinor space
of protons and neutrons separately; the upper and lower spinor indices
refer to protons and neutrons, respectively. The first indices, $b$ and $\alpha$,  form
an outer product of two columns, whereas the second ones, $c$ and $\beta$,
form an outer product of two rows. To explicitly establish a
relation of the amplitude (\ref{eq5}) with
the corresponding amplitude (\ref{ampl}) of the $NN\to NNV$ process we
find the spinor structure by a decomposition of the operator $\hat O$ 
in each of its indices over the corresponding complete set of Dirac spinors, i.e,
\be
\hspace*{-1.5cm}
\hat O^{bc}_{\alpha\beta}(12;1'2',\CM_V)=\frac{1}{(2m)^4}\sum\limits_{r,r',\rho,\rho'=1}^4\,
A_{r,r';\rho,\rho'}^{\CM_V}(12;1'2')
 u^b_{r'}(1') \bar u^c_{r}(1)  \bar u^\beta_{\rho}(2)  u^\alpha_{\rho'}(2'), \label{eq6} 
\ee
where the coefficient
$A_{r,r';\rho,\rho'}^{\CM_V}(12;1'2')$ is determined by the
completeness and orthogonality of the Dirac spinors, $\bar
u_r({\bf p}) u_{r'}({\bf p}) = \varepsilon_r 2m\delta_{rr'}$,
yielding
\be
 A_{r,r';\rho,\rho'}^{\CM_V}(12;1'2')=\varepsilon_r\varepsilon_{r'}\varepsilon_\rho
\varepsilon_{\rho'} \bar u^b_{r'}(1')\, \bar
u^\alpha_{\rho'}(2')\, \hat O^{bc}_{\alpha\beta}(12;1'2',\CM_V)
u^c_{r}(1)\,u^\beta_{\rho}(2), \label{eq7}
\ee
where
$\varepsilon=+1$ for  $r=1,2$ and $\varepsilon=-1$ for $r=3,4$.
Since the defined  operator (\ref{eq6})
generally acts in the nucleon spinor space
its matrix elements describe processes with anti-particles as well.
This is explicitly reflected  in the dependence of
the coefficients  $A_{r,r';\rho,\rho'}^{\CM_V}(12;1'2')$
on spinors with anti-particle quantum numbers. In our case
this dependence can occur in the process (\ref{reaction})
only as virtual creation and annihilation of $N\bar N$ pairs, which
within the BS formalism are allowed through the presence of
negative-energy $P$ components in the BS amplitude.
However, in the considered kinematics the contribution of $P$ waves
is by far smaller than the main $S^{++}$ and $D^{++}$ components~\cite{quad}
and consequently, in what follows we disregard all the partial BS amplitudes
with at least one negative energy index, leaving only
the  $S^{++}$ and $D^{++}$ waves.
Then substituting (\ref{eq6}) into (\ref{eq5}) one obtains
\be
T_{s_1s_2}^{\CM_V\CM_d}(s,t)=\frac{-i}{(2m)^2}\int\,\frac{d^4p}{(2\pi)^4}
\bar \Phi_{\CM_d}^{\alpha b}(1',2')\, \sum\limits_{rr'=1}^2
A_{s_1s_2;rr'}^{\CM_V}(12;1'2')
u^\alpha_{r'}(2')u^b_r(1')\nonumber\\ &&
=\frac{i}{(2m)^2}\sum\limits_{rr'=1}^2 \int\,\frac{d^4p}{(2\pi)^4}
\left( u^\alpha_{r'}(2')\right )^T\gamma_c^{\alpha\alpha'}\left(
\gamma_c^{\alpha'\alpha''} \bar \Phi_{\CM_d}^{\alpha''b}(1',2')\right )\,
 A_{s_1s_2;rr'}^{\CM_V}(12;1'2')
u^b_r(1')\nonumber\\&&
 = \frac{i}{(2m)^2}\sum\limits_{rr'=1}^2
\int\,\frac{d^4p}{(2\pi)^4} A_{s_1s_2;rr'}^{\CM_V}(12;1'2') \bar { v}_{r'}(2')
\bar\Psi_{\CM_d}(1',2')u_r(1'), \label{eq8} 
\ee 
where $\gamma_c$ is the charge conjugation matrix,
$\bar v_r(2')\equiv u^T_r(2')\gamma_c$, and the new BS amplitude $\bar \Psi_{\CM_d}(1',2')\equiv
\gamma_c\bar\Phi_{\CM_d}(1',2')$  is written now in the form of a $4\otimes 4$ matrix and
represents the solution of the BS equation written also in  matrix form. Note that in
eq.~(\ref{eq8}) the spinor $\bar { v}_{r'}(2')$ does not describe an anti-particle, 
it is rather a consequence of the efforts made to cast the BS amplitude in a matrix form 
(see for details, e.g., refs.~\cite{Tjon}) and still refers to nucleons.

\begin{figure}[h]  %      Fig10
\includegraphics[width=1.0\textwidth]{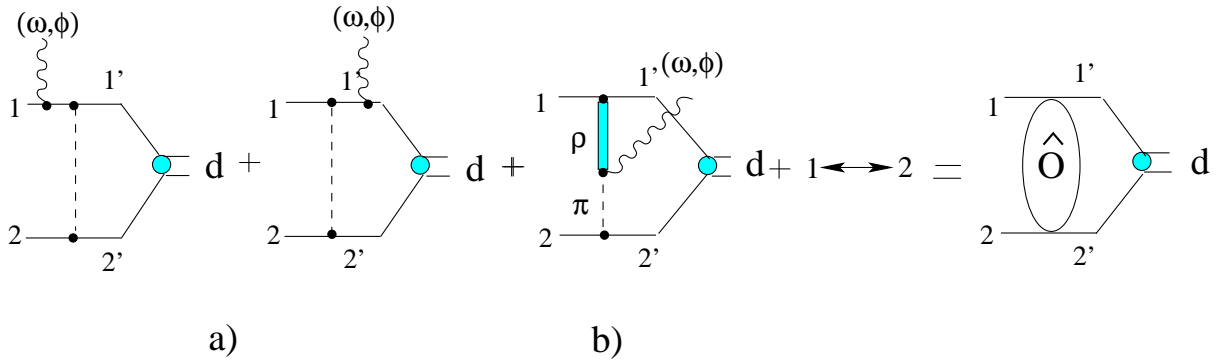} %
\caption{Diagrams contributing to the
process $pn\to dV$. The sum of bremsstrahlung (a) and conversion-type (b)
diagrams results in the matrix element of the operator $\hat O(12;1'2'V)$ (fig.~\ref{pict1})
sandwiched between the initial two-nucleon states and the final deuteron.}
\label{pict10}
\end{figure}

As in the previous case of $NN$ reactions, the effective interactions in eq.~(\ref{eq9}) again
result into two types of diagrams, conversion currents and bremsstrahlung emission, which are
depicted in fig.~\ref{pict10}. Consequently, having already computed the operator $\hat O$, it is
straightforward to obtain the coefficients $A_{r,r';\rho,\rho'}^{M_V}(12;1'2')$ in eq.~(\ref{eq7})
within the OBE approximation. Note that in case when all particles are on mass shell,
$A_{r,r';\rho,\rho'}^{M_V}(12;1'2')$ exactly coincides with the amplitude (\ref{ampl}) of the
elementary process (\ref{reac1}), i.e.,
$A_{r,r';\rho,\rho'}^{M_V}(12;1'2')=T_{r,r';\rho,\rho'}^{\CM_V }(12;1'2'V)$.
However, in general  this amplitude corresponds to
a virtual process of vector meson production
with two off-mass shell nucleons in the final state.

Since our numerical solution for the BS equation has been obtained in the
deuteron center of mass \cite{solution}, all further calculations will be performed 
in this system. First
we introduce the relevant kinematical variables as follows:
$p_1$ and $p_2$ are the four momenta of incoming nucleons, $p_1'$ and $p_2'$
stand for the  momenta
of the internal (off-mass shell) nucleons in the deuteron with  $p=(p_1'-p_2')/2$,
$\xi_\CM$ denotes the polarization 4-vectors of the deuteron and vector mesons.
In this notation the BS amplitudes in the deuteron  rest system are of the
form \cite{quad}
\begin{eqnarray}
&&
\!\!\!\!\!
\Psi_{\CM_d}^{S^{++}}(p_1',p_2')={\cal N}(\hat k_1+m )\frac{1+\gamma_0}{2}\hat\xi_{\CM_d}(\hat k_2-m)
\phi_S (p_0,|{\bf p}|), \label{psis}\\[2mm]
&&\!\!\!\!\!
\Psi_{\CM_d}^{D^{++}}(p_1',p_2')=-\frac{{\cal N}}{\sqrt{2}}
(\hat k_1+m )\frac{1+\gamma_0}{2}
\left (
\hat\xi_{\CM_d} +\frac{3}{2|{\bf p}|^2} (\hat k_1-\hat k_2)(p\xi_M)\right )
(\hat k_2-m)
\phi_D (p_0,|{\bf p}|),\nonumber
%\label{psid},
\end{eqnarray}
where
$k_{1,2}$ are on-mass shell 4-vectors related  to the
off-mass shell vectors $p_{1,2}'$ as follows
\begin{equation}
k_1=(E_{\bf p},{\bf p}),\quad k_2=(E_{\bf p},-{\bf p}),\quad p_1'=(p_{10}',{\bf p}),\quad
p_2'=(p_{20}',-{\bf p}),\quad E_{\bf p}=\sqrt{{\bf p}^2+m^2},
\nonumber
\end{equation}
and $\phi_{S,D} (p_0,|{\bf p}|)$ are the partial scalar amplitudes, related to
the corresponding partial vertices as
\begin{equation}
\phi_{S,D} (p_0,|{\bf p}|)=
\displaystyle\frac{G_{S,D} (p_0,|{\bf p}|)}{\left(\displaystyle\frac{M_D}{2}-E_{\bf p}\right)^2-p_0^2}.
\nonumber
\end{equation}
$M_d$ is the deuteron mass, and the normalization factor reads 
${\cal N}=\displaystyle\frac{1}{\sqrt{8\pi}}\displaystyle\frac{1}{2E_{\bf p}(E_{\bf p}+m)}$. 
The components of the polarization  vector of a vector particle moving with 4-momentum  
$p=(E,{\bf p})$, having the  polarization projection $\CM=\pm 1,\, 0$ and mass $M$ are
\begin{eqnarray}
&&
\xi_\CM=\left( \frac {\bf p \boldxi_\CM}{M},
\boldxi_\CM+{\bf p}
\frac {\bf p \boldxi_\CM}{M(E_{\bf p}+M)} \right ),
\label{xi}
\end{eqnarray}
where $\boldxi_\CM$ is the polarization vector for the particle at rest,
\begin{equation}
\begin{array}{ccc}
\boldxi_{+ 1} =-\displaystyle\frac{1}{\sqrt{2}}\,
\left ( \begin{array}{c}1\\ i\\0 \end{array} \right ),
&
\quad \boldxi_{- 1}=\displaystyle\frac{1}{\sqrt{2}}
\left ( \begin{array}{c}1\\ -i\\0 \end{array} \right ),
&
\quad \boldxi_{0}=
\left ( \begin{array}{c} 0\\ 0\\1 \end{array} \right ). \label{xilab}
\end{array}
\end{equation}
The Dirac spinors, normalized as $\bar{u}(p) u(p)=2m$ and $\bar{v}(p) v(p)=-2m$, read
\begin{eqnarray}
u({\bf p},s)=\sqrt{m+ E_{{\bf p}}}
\left ( \begin{array}{c} \chi_s \\ \displaystyle\frac{\boldsigma {\bf p}}{m+ E_{{\bf p}}}\chi_s
\end{array}  \right), \quad \quad
v({\bf p}\,,s)=\sqrt{m+ E_{{\bf p}}}\left( \begin{array}{c}
\displaystyle\frac{\boldsigma {\bf p}}{m+ E_{{\bf p}}}\widetilde\chi_{s}\\
 \widetilde\chi_{s}
\end{array}  \right), \label{spinors}
\end{eqnarray}
where $\widetilde\chi_s\equiv -i\sigma_y\chi_s$,  and $\chi_s$ denotes the usual two-dimensional
Pauli spinor. In general, the BS amplitude consists on eight partial components. As already
mentioned, in eq.~(\ref{psis}) we take into account only the most important ones, namely the $S$
and $D$ partial amplitudes. The other six amplitudes may become important at high transferred
momenta \cite{ourphysrev,quad}, hence for the present near-threshold process (\ref{reaction}) they
may be safely disregarded. Substituting eqs. (\ref{psis})-(\ref{spinors}) into (\ref{eq8}) one
obtains after some algebra \cite{ourPhi} 
\be \!\!\!\!\!\!\!\!
T_{s_1s_2}^{\CM_V\CM_d}(s,t)=\frac{-i}{\sqrt{8\pi}}\sqrt{|\CM_d|+1}\nonumber\\&&
\!\!\!\!\!\!\!\!\times \sum\limits_{rr'} \int\,\frac{d^4p}{2E_{\bf p}(2\pi)^4}
\frac{G_S-G_D/\sqrt{2}}{\left(M_D/2-E_{\bf p}\right)^2-p_0^2}A_{s_p,s_n;rr'}^{\CM_V} ({\bf
p}_1,{\bf p}_2;{\bf p}_1',{\bf p}_2',{\bf
P}_V)\delta_{r+r';\CM_d}\label{eq15}\\&&\!\!\!\!\!\!\!\!+ \frac{3i}{\sqrt{16\pi}}\sum\limits_{rr'}
\int\,\frac{d^4p}{2E_{\bf p}(2\pi)^4} \frac{G_D}{\left(M_D/2-E_{\bf p}\right)^2-p_0^2}
A_{s_p,s_n;rr'}^{\CM_V}({\bf p}_1,{\bf p}_2;{\bf p}_1'{\bf p}_2'{\bf P}_\phi)
\widetilde\chi^+_{r'}\, (\boldsigma\bn)\chi_r\,(\bn\boldxi^*_{\CM_d}).\nonumber 
\ee 
By closing the integration contour in the upper hemisphere and 
picking up the residuum in $p_0 = M_d/2 - E_{\bf p}$
and introducing the notion of the deuteron $S$ and $D$ wave functions as 
\be
u_S(p)=\frac{G_S(p_0,|{\bf p}|)/4\pi}{\sqrt{2M_D}(2E_{\bf p}-M_d)};\quad
u_D(p)=\frac{G_D(p_0,|{\bf p}|)/4\pi}{\sqrt{2M_D}(2E_{\bf p}-M_d)}; \nonumber\\&& {\rm with}
\qquad \frac{2}{\pi}\int |{\bf p} |^2 \,d\,|{\bf p}| (u_S^2+u_D^2)\approx 1, \label{norm} 
\ee 
the final expression for the amplitude $T$ may be cast in the form 
\be
T_{s_1s_2}^{\CM_V\CM_d}(s,t)=\sqrt{\frac{M_D}{4\pi}} \sum\limits_{rr'} \int\,\frac{d^3{\bf
p}}{E_p(2\pi)^2} A_{s_p,s_n;rr'}^{\CM_V}({\bf p}_1,{\bf p}_2;{\bf p},-{\bf p},{\bf P}_V)
\nonumber\\&&\times \left\{ \sqrt{|\CM_d|+1} \left[u_S(p)-\frac{u_D(p)}{\sqrt{2}}\right
]\delta_{r+r';\CM_d}- 3\frac{u_D(p)}{\sqrt{2}}\left(\boldxi^*_{\CM_d}\bn\right)\,\widetilde
\chi_{r'} (\boldsigma\bn)\chi_r\right\}, \label{finalT} 
\ee
where $\bn$ is a unit vector parallel to ${\bf p}$.
Equation (\ref{finalT}) may be written in
a more familiar form as to better emphasize the relation of our formulae with
their non-relativistic analogues. For this sake observe that in eq.~(\ref{finalT}) one has
$\sqrt{|\CM_d|+1}\delta_{r+r';\CM_d}
= - \sqrt{2}\langle \frac 12\ r \frac12 r'|1\CM_d\rangle$ and
$\left(\boldxi^*_{\CM_d}\bn\right)\,(\boldsigma\bn)=(4\pi/3)\sum\limits_{\alpha,\beta}
(-1)^{\alpha+\beta}  \boldxi^*_{-\alpha}\boldsigma_{-\beta} {\rm Y}_{1\alpha}(\bn)
{\rm Y}_{1\beta}(\bn) $, 
where ${\rm Y_{lm}}(\bn)$ are the usual spherical harmonics.
Then, by making use of the addition theorem for the spherical harmonics
$ {\rm Y_{lm}}(\bn)$
and the Wigner-Eckart theorem for the matrix elements between states
with definite angular momenta, the amplitude
$T_{s_1s_2}^{\CM_V\CM_d}(s,t)$ becomes
\begin{eqnarray}  &&\hspace*{-0.1cm}
T_{s_1s_2}^{\CM_V\CM_d}(s,t)=\sqrt{2M_D}\sum\limits_{rr'}
\int\,\frac{d^3{\bf p}}{E_p(2\pi)^2}
A_{s_p,s_n;rr'}^{\CM_V}({\bf p}_1,{\bf p}_2;{\bf p},-{\bf p},{\bf P}_V)
\label{newT} \times\\ &&
\left\{   \left\langle \frac12 r\frac12 r'|1\CM_d\right\rangle  {\rm Y_{00}}(\bn) u_S(p)-
\sum\limits_{{\rm m},\nu_{12}} \left\langle 2{\rm m} 1 \nu_{12}|1\CM_d\right\rangle
\left\langle \frac12 r\frac12 r'|1\nu_{12} \right\rangle {\rm Y_{2m}^*}(\bn)  u_D(p)
\right\}, \nonumber
\end{eqnarray}
where the last line exactly coincides  with a
non-relativistic spin overlap between the deuteron wave function and two
Pauli spinors of intermediate nucleons. Note that usually in non-relativistic 
meson-nucleon theories the analogue of the amplitude
$T_{s_1s_2}^{\CM_V\CM_d}(s,t)$ is obtained by a non-relativistic reduction
of the initially covariant
operator $\hat O$ with subsequent calculations of the matrix element
\begin{eqnarray} &&
T_{s_1s_2}^{\CM_V\CM_d \ N.R.}(s,t)=
\left\langle D,\CM_d \left |\, \hat O^{N.R.}(12;1'2'V)\,\right |{\bf p}_1,s_1,{\bf p}_2,s_2
\right\rangle \nonumber\\ && =
\sum\limits_{r,r'=1}^2 \left\langle D,\CM_d\ |k_1,r,k_2,r'\right\rangle
\langle k_1,r,k_2,r'\
|\hat O^{N.R.}(12;1'2'V)\ |{\bf p}_1,s_1,{\bf p}_2,s_2 \rangle, \label{nr}
\end{eqnarray}
where now $\left\langle D,\CM_d\ |k_1,r,k_2,r'\right\rangle$ is
the non-relativistic (complex conjugated)
deuteron wave function in the momentum space.
In our case, the use of the BS formalism allows to compute the matrix element (\ref{newT})
directly from the covariant expression of the previously  found
operator $\hat O$, avoiding the cumbersome procedure of
its non-relativistic reduction for non-relativistic calculations by eq.~(\ref{nr}).
However, since in the considered kinematical range
the BS wave functions, eq.~(\ref{norm}), are rather similar to their non-relativistic
analogues (see, e.g.\ \cite{quad}) and due to the formal equivalence of
eqs.~(\ref{newT}) and (\ref{nr}) we can use the former one
for non-relativistic calculations as well
by merely replacing  $u_S(p)$ and $u_D(p)$
with the corresponding quantities from the
Schr\"odinger equation with, e.g., Bonn or Paris potentials.

\subsection{Results for $\mbox{\boldmath$pn \to d V$}$}\label{subDeu2}

In our calculation of the cross section of the process (\ref{reaction})
we use the numerical solution of the BS equation \cite{solution} in  ladder
approximation obtained with a realistic OBE interaction.
The effective parameters used in the ladder approximation
when solving the BS equation have been fixed in such a way  to obtain
a good description of the $NN$ elastic scattering data and the
main static properties of the deuteron \cite{quad}. The obtained
parameters  turn out to be very close to those obtained in the
non-relativistic framework of the Bonn group in determining the
OBE $NN$ potential~\cite{bonncd}. In this sense our analysis
of the processes with the deuteron is consistent
with the previous consideration of $NN\to NNV$ processes.

\begin{figure}[h]  %      Fig11
\includegraphics[width=0.95\textwidth]{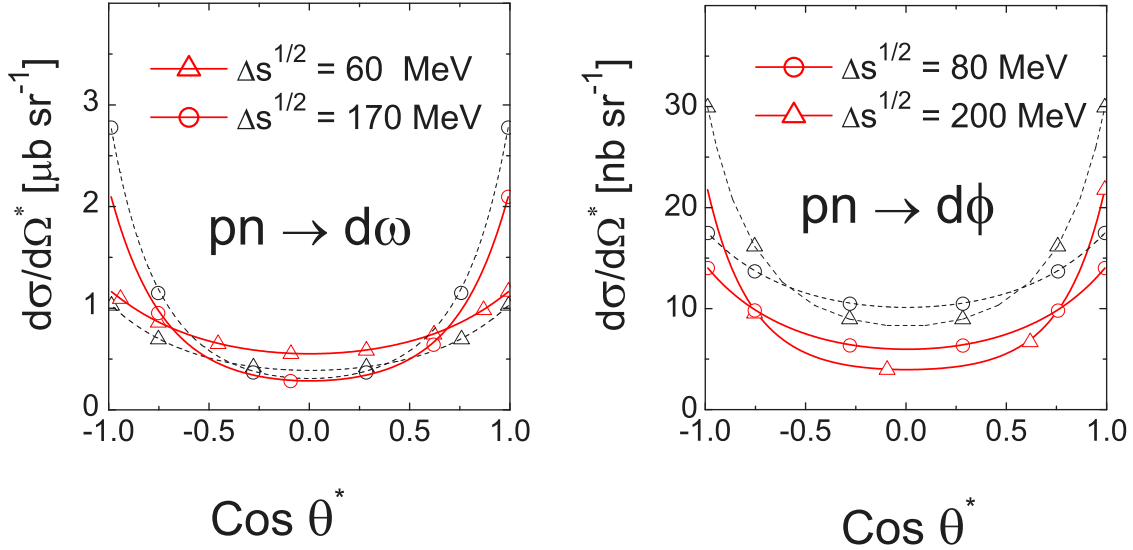} %
\caption{Angular distributions for $\omega$ (left panel) and $\phi$ (right panel)
mesons produced in $pn\to dV$ processes for various values of
the excess energy $\Delta s^{1/2}=\sqrt{s}-M_d-m_V$ ($V=\omega,\phi$).
Solid lines: results of calculations with the deuteron wave function
obtained within the BS formalism; dashed lines: the non-relativistic
deuteron wave function with  the Bonn potential.
The curves labelled with open circles correspond to values of
$\Delta s^{1/2}$ for which the experimental data of elementary process $pp$ exits (cf.\
figs.~\ref{pict2}-\ref{pict6}); the curves labelled with triangles are presented for an
illustration of  the change of the distributions with the excess energy.} \label{pict11}
\end{figure}

\begin{figure}[h]  %      Fig12
\includegraphics[width=0.85\textwidth]{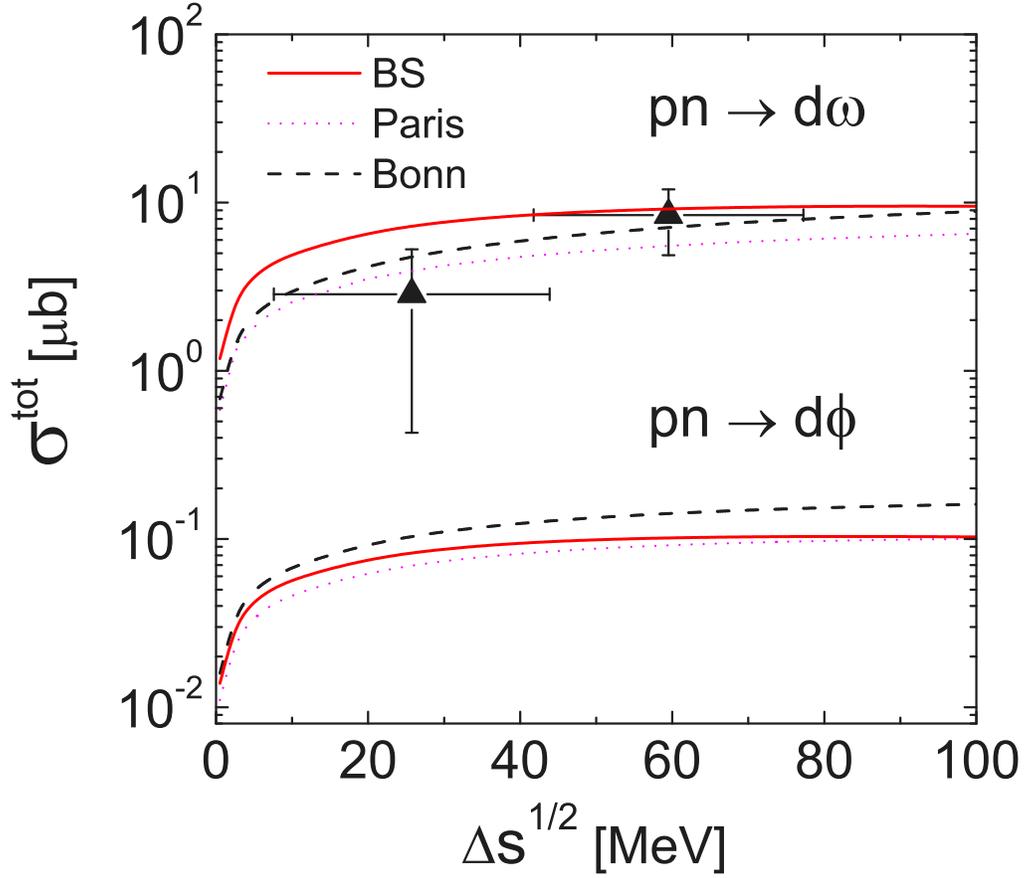} %
\caption{Total cross sections  for  $\omega$  and $\phi$
mesons for $pn\to dV$ reactions as a function
the excess energy $\Delta s^{\frac12}=\sqrt{s}-M_d-m_V$ ($V=\omega,\phi$).
Solid lines: results of calculations with the deuteron wave function
obtained within the BS formalism; dashed (dotted) lines: using the non-relativistic
deuteron wave function with the Bonn (Paris) potential.
The experimental data are from COSY-ANKE \cite{ankeExper}.}
\label{pict12}
\end{figure}

In figs.~\ref{pict11} and \ref{pict12} the angular distribution and the total cross section
are depicted. The shape of our evaluated cross section is rather similar to one 
computed in ref.~\cite{nakayama}. However,
there is some difference in the absolute values in case of $\phi$. This
difference is probably due to the fact that we fitted our parameters to $\pi N\to \phi N$
data and to $N N \to \phi NN$ data, in particular, the new DISTO data point with 
$\sigma=0.17\ \mu b$ \cite{DISTO2}. In addition, the methods of calculating the
FSI effects are slightly different, which could provide some
difference in  values  of FSI corrections as well. Nevertheless, the magnitude of our parameters
and those of ref.~\cite{nakayama} are very close to each other. 
It is also worth emphasizing that,
as discussed in refs.~\cite{titov,nakayama}, there can be several sets of parameters equally well
describing the $pp\to ppV$ data. These sets differ not only by absolute  values of parameters but
also by the relative contributions of bremsstrahlung and conversion diagrams which possess
different isospin structure and, hence differently contribute to $pp$ and $pn$ processes. So, in
case of $pn\to d V$ processes the isospin transition corresponds to $\Delta {\cal T}=0$,
consequently the conversion diagrams are enhanced by a factor of "-3" in comparison with the
bremsstrahlung diagram with exchange of neutral mesons.
That means that
(i) the contributions of bremsstrahlung terms
are strongly reduced in $pn \to dV$ processes
in comparison with the elementary $pp$ reaction
and (ii) the shape of the cross sections and angular distributions in the
process $pn\to dV$ follows the behavior of the corresponding
distribution  in the elementary processes, modified by the deuteron wave function. This is clearly
seen in fig.~\ref{pict11}, where the shape of the angular distribution is very similar to the
distributions found in $pn$ reactions (the shapes of the corresponding  distributions in
figs.~\ref{pict2} and~\ref{pict5} must be compared  with the curves labelled by open circles in
fig.~\ref{pict11}).
At the threshold the distribution is fairly flat, while with increasing energy
some structure around the forward-backward
directions becomes pronounced. The shape of the differential
cross section in near forward and backward directions depends
essentially upon the parameter set used in calculations (for a detailed
analysis of the dependence of the shape of the angular distributions
upon the chosen parameters, see \cite{nakayama1,nakayama2}).

In fig.~\ref{pict12} the total cross sections for $\omega$ and $\phi$ 
meson production are depicted as a function of the excess energy. 
The $\omega$ production is larger than the $\phi$ cross section by roughly two orders of
magnitude. From figs.~\ref{pict11} and \ref{pict12} we observe that
the computed cross sections exhibit a dependence on the potential used in the calculations of the
deuteron wave function. This dependence is more pronounced at relatively low energies and has
opposite behavior in $\omega$ (the BS wave function provides slightly larger cross sections
than the Bonn one) and $\phi$ production (the BS cross section is smaller
than the Bonn one). This difference decreases with increasing energy. This
can be observed from a comparison of angular distributions at two
different energies (compare the curves labelled with open circles with the ones labelled by
triangles in  fig.~\ref{pict11}). In fig.~\ref{pict12} this effect is seen for $\omega$ production
(the upper panel) and not yet visible for $\phi$ mesons (lower panel).

\begin{figure}[h]  %      Fig13
\includegraphics[width=0.85\textwidth]{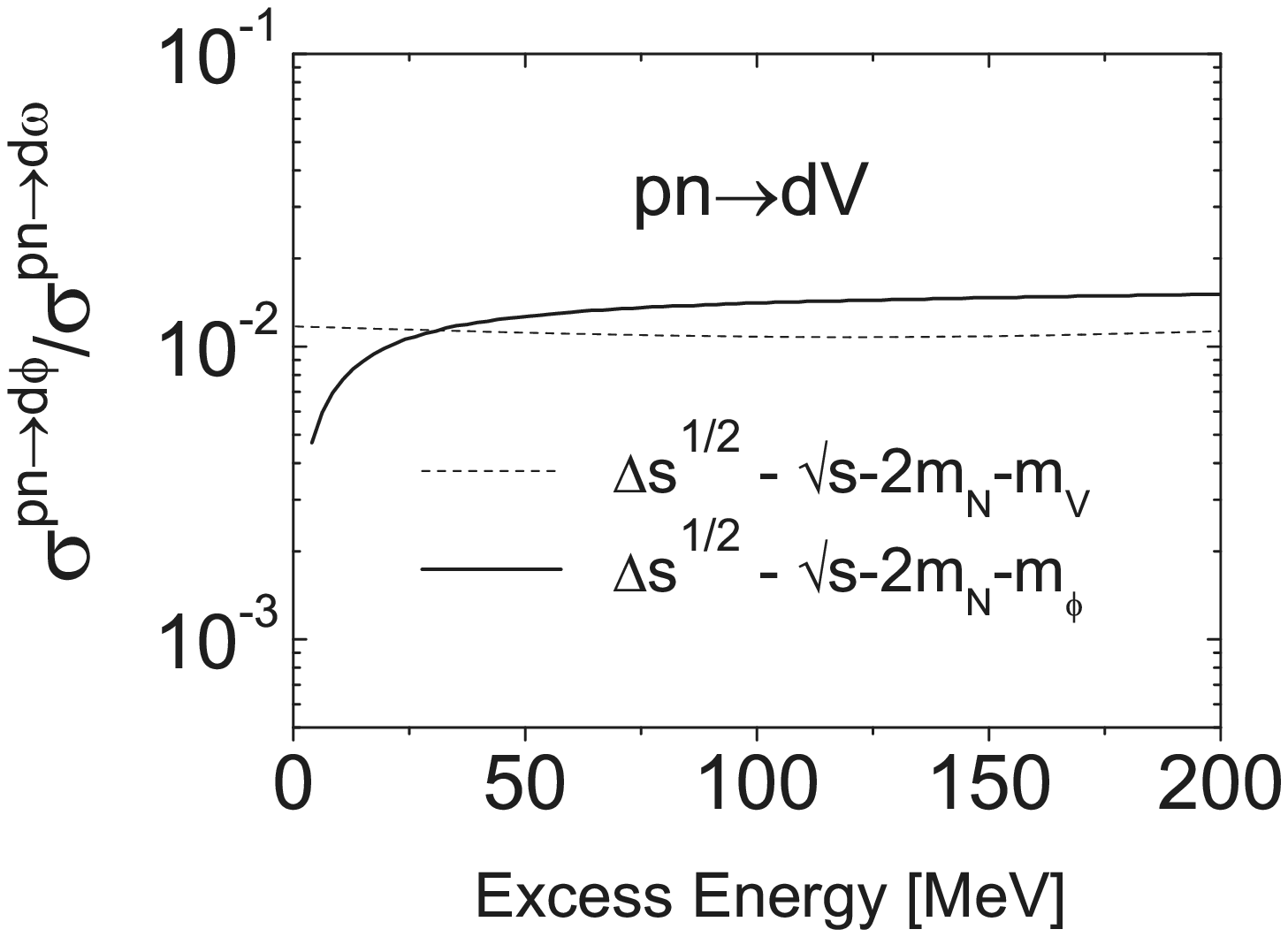} %
\caption{The OZI rule ratio for  $pn\to dV$ processes as a function of the excess energy.
The dashed line represents the ratio at equal values of the excess energy,
while the solid line reflects the ratio at equal beam energies and is measured from the
threshold of $\phi$ meson production.} \label{pict13}
\end{figure}

Eventually, in fig.~\ref{pict13} we analyze the values of the OZI ratio defined at equal excess
energies (dashed line) and at equal beam energies (solid line). 
Corresponding to our approach in
both cases the FSI effects are the same (i.e., the deuteron in the final state). The only
difference should appear due to the difference in the phase space volumes for
$\omega$ and $\phi$ mesons which
is estimated to be minimized for the ratio at equal excess energy near the
threshold and more pronounced for the ratio at equal beam energies.
As the energy increases the difference in the two definitions should not be
too large, as seen in fig.~\ref{pict13}. In both cases the OZI ratio is essentially
larger than the one expected form the naive OZI rule restrictions.

\section{Summary}\label{sumary}

In summary we present a combined analysis of  vector meson
production in  $pp\to ppV$,
$pn\to pnV$ and $pn\to dV$ processes for $V=\omega, \phi$. The elementary reactions
$NN$ are treated within an effective meson-nucleon theory with
most parameters fixed from independent experiments.
The few remaining parameters are fitted to achieve a reasonably good description
of the available angular distributions. With these parameters at our
disposal we obtain a fairly
good description of the energy dependence of total cross sections.
Then the OZI rule is analyzed for different definitions of the relevant ratio of $\phi/\omega$
meson production. It is found that an enhancement of the OZI rule  ratios can be obtained without
any expected violation of the original rule and could be interpreted as dynamical effect which
occurs as a sophisticated interference of different types of diagrams and isospin effects. Using
the same meson-nucleon theory we investigate vector production with a deuteron in the final
state. In order to be able to use directly the results from elementary reaction we apply the
Bethe-Salpeter formalism for the deuteron in the final state. The final expressions are presented
in a form fully corresponding to the non-relativistic approach. Calculations with the same set of
parameters show that within the proposed approach one can obtain reasonable values of the total
cross section, close to the preliminary experimental data
from ANKE. The OZI rule ratio is found to be almost independent
of the energy and, similarly to the $NN$ case, is enhanced relative
to the naive expectations, based on the OZI rule restrictions.

The proposed approach allows to calculate a number of polarization
observables in the considered processes which could be directly related with the OZI
rule. However this task is beyond of the goal of the present paper.
An analysis of polarization observables in $NN\to NN\phi$ reactions can be
found in ref.~\cite{titov} and a for $pn\to d\phi$ in ref.~\cite{ourPhi}.

Our approach extends the study of \cite{nakayama2} by including the neutron channels and the
deuteron final state. First experimental data for $pn\to d\omega$ is at our disposal, while for
$pn\to d\phi$ data are expected soon from COSY-ANKE. It should be noted that in \cite{nakayama2}
also a few nucleon resonances are taken into account for $\omega$. 
This leads to a renormalization of
parameters, coupling constants and cut-off's, and invokes further new parameters. 
The role of baryon resonances is, in particular, stressed in \cite{Fuchs2}.
The comprehensive extension of our approach with inclusion 
of the full list of resonances deserves a separate investigation, 
e.g.\ along the lines of \cite{Mosel,M.Lutz,Sasha}.

\subsection*{Acknowledgements}
We thank  H.W, Barz, A.I. Titov, S.S. Semikh for useful discussions.
L.P.K. would like to thank for the warm hospitality in the Research Center Rossendorf.
This work has been supported by the BMBF grant 06DR121.

\newpage
%%%%%%%%%%%%%%%%%%%%%%%%%%%%%%%%%%%%%%%%%% BEGIN FIGURES  %%%%%%%%%%%%%%%%%

%%%%%%%%%%%%%%%%%%%%%%%%%%%%%%%%%%%%%%%%%%  Fig.2

%%%%%%%%%%%%%%%%%%%%%%%%%%%%%%%%%%%%%%%%%% Fig 3.  %%%%%%%%%%%%%%%%%%%%%%%%%%%%%%%%%%%%%

%%%%%%%%%%%%%%%%%%%%%%%%%%%%%%%%%%%%%%%%%%%% Fig.4

%%%%%%%%%%%%%%%%%%%%%%%%%%%%%%%%%  Fig.5  %%%%%%%%%%%%%%%%%%%%%%%%%%%%%%%%%%%%%%%%%%%

%%%%%%%%%%%%%%%%%%%%%%%%%%%%%%%%%     Fig.6   %%%%%%%%%%%%%%%%%%%%%%%%%%%%%%%%%%%%%%%%%%%%%%

%%%%%%%%%%%%%%%%%%%%%%%%%%%%%% Fig7 %%%%%%%%%%%%%%%%%%%%%%%%%%%%%%%%%%%%5

%%%%%%%%%%%%%%%%%%%%%%%%%%%%%%%%         Fig.8.    %%%%%%%%%%%%%%%%%%%%%%%%%%%%%%%%%%%%%%%%%%%%%%

%%%%%%%%%%%%%%%%%%%%%%%%%%%%%          Fig.9   %%%%%%%%%%%%%%%%%%%%%%%%%%%%%%%%%%%%%%%%%%5

%%%%%%%%%%%%%%%%%%%%%%%%%%%%%%%%%%%%  Fig.10 %%%%%%%%%%%%%%%%%%%%%%%%%%%%%%%%%%%%%%%%%%

%%%%%%%%%%%%%%%%%%%%%%%%%%%%%%%  Fig. 11  %%%%%%%%%%%%%%%%%%%%%%%%%%%%%%%%%%%%%%%%%5

%%%%%%%%%%%%%%%%%%%%%%%%%%%%%%  Fig 12%%%%%%%%%%%%%%%%%%%%%%%%%%%%%%%%%%%%%%%%%

%%%%%%%%%%%%%%%%%%%%%%%%%%%%  Fig.13   %%%%%%%%%%%%%%%%%%%%%%%%%%%%%%%%%%%%%%%%%%%%%%

%%%%%%%%%%                  END   DOCUMENT

\begin{thebibliography}{99}
\bibitem{ozi}
S. Okubo, Phys. Lett. {\bf 5} (1963) 165;\\
G. Zweig, CERN Report 8419/TH 412 (1964);\\
I. Iizuka, Prog. Theor. Phys. Suppl. {\bf 37/38} (1966) 21.
\bibitem{sakurai}
J.J. Sakurai, Phys. Rev. Lett. {\bf 9} (1962) 472.
\bibitem{gelmann}
M. Gell-Mann, Phys. Rev. {\bf 125} (1962) 1067.
\bibitem{okubo77}
S. Okubo, Phys. Rev. {\bf D 16} (1977) 2336.
\bibitem{DISTO1}
F. Balestra at al. (DISTO Collaboration), Phys. Rev. Lett. {\bf 81} (1998) 4572.
\bibitem{DISTO2}
F. Balestra at al. (DISTO Collaboration) Phys. Rev. {\bf C 63} (2001) 024004;
Phys. Lett. {\bf B 468} (1999) 7.
\bibitem{johansson}
R. Bilger et al., Nucl. Inst. Meth. {\bf A 457} (2001) 64;\\
H. Calen et al., Phys. Rev. Lett. {\bf 80} (1998) 2069.
%T. Johansson, B. H\"oistad, TSL proposal, unpublished.
\bibitem{anke}
M. B\"uscher et al., "Study of $\omega$ and $\phi$ meson
production in the reaction $pD\to DVp_{sp}$ at ANKE, Exp. \#75/ANKE.
\bibitem{sibirtsev}
A. Sibirtsev, W. Cassing, Eur. Phys. J. {\bf A 7}  (2000) 407.
\bibitem{ellis}
J.R. Ellis, M. Karliner, D.E. Kharzeev, M.G. Sapozhnikov,
Nucl. Phys. {\bf A 673} (2000) 256;
%J.R. Ellis, M. Karliner, D.E. Kharzeev, M.G. Sapozhnikov,
Phys. Lett. {\bf B 353} (1995) 319.
\bibitem{rotz}
V.E. Markushin, M.P. Locher, Eur. Phys. J. {\bf A 1} (1998) 91;\\
S. von Rotz, M.P. Locher, V.E. Markushin, hep-ph/9912359.
\bibitem{donoghue}
J.F. Donoghue, C.R. Nappi, Phys. Lett. {\bf B 168}  (1986) 105.
\bibitem{gasser}
J. Gasser, H. Leutwyler, M.E. Sainio, Phys. Lett. {\bf B 253} (1991) 252.
\bibitem{ashman}
J. Ashman et al., Phys. Lett. {\bf B 206} (1988) 364.
\bibitem{isgur}
N. Isgur, H.B. Thacker, hep-lat/0005006 (2000); Phys. Rev. {\bf D 64}  (2001) 094507.
\bibitem{shuryak}
T. Sch\"afer, E.V. Shuryak, hep-lat/0005025.
\bibitem{golubeva}
L.A. Kondratyuk, Ye. Golubeva, M. B\"uscher, nucl-th/9808050.
\bibitem{grishina}
V.Yu. Grishina, L.A. Kondratyuk, M. B\"uscher, nucl-th/9906064.
\bibitem{nakayama}
K. Nakayama, J. Haidenbauer, J. Speth, Phys. Rev. {\bf C 63} (2001) 015201.
\bibitem{nakayama1}
K. Nakayama, J.W. Durso, J. Haidenbauer, C. Hanhart,  J. Speth, 
Phys. Rev. {\bf C 60} (1999) 055209.
\bibitem{ourPhi}
L.P. Kaptari, B, K\"ampfer, Eur. Phys. J. {\bf A 14} (2002) 211.
\bibitem{nakayama2}
K. Tsushima, K. Nakayama, Phys. Rev. {\bf C 68} (2003) 034612.
\bibitem{Fuchs} A. Faessler, C. Fuchs, M.I. Krivoruchenko, B.V. Martemyanov,
Phys. Rev. {\bf C 68} (2003) 068201.
\bibitem{Tjon}
B.D. Keister, J.A. Tjon, Phys. Rev. {\bf C 26} (1982) 578;\\
G. Rupp, J.A. Tjon, Phys. Rev. {\bf C 41} (1990) 472;\\
J.J. Kubis, Phys. Rev. {\bf D 6} (1972) 547;\\
M.J. Zuilhof, J.A. Tjon, Phys. Rev. {\bf C 22} (1980) 2369.
\bibitem{solution}
A.Yu. Umnikov, L.P. Kaptari, F.C Khanna,
Phys. Rev. {\bf C 56} (1997) 1700;\\
A.Yu. Umnikov, L.P. Kaptari, K.Yu. Kazakov, F.C. Khanna,
Phys. Lett. {\bf B 334} (1994) 163;\\
A.Yu. Umnikov, Z. Phys. {\bf A 357} (1997) 333.
\bibitem{bonncd}
R. Machleidt, Adv. Nucl. Phys. {\bf 19} (1989) 189; Phys. Rev. {\bf C 63} (2001) 024001.
\bibitem{chung}
W.S. Chung, G.Q. Li, C.M. Ko, Phys. Lett. {\bf B 401} (1997) 1.
\bibitem{titov}
A.I. Titov, B. K\"ampfer, B.L. Reznik, Eur. Phys. J., {\bf A 7} (2000) 543;
Phys. Rev. {\bf C 65} (2002) 065202.
\bibitem{durso}
J.W. Durso, Phys. Lett. {\bf B 184} (1987) 348.
\bibitem{dashen}
R.D. Dashen, D.H. Sharp, Phys. Rev. {\bf 133} (1964) B1585.
\bibitem{TOF}
S.Abd El-Salam et al. (COSY-TOF),  Phys. Lett. {\bf B 522} (2001) 16.
\bibitem{Hibou} 
F. Hibou et al. (SATURNE), Phys. Rev. Lett. {\bf 83} (1999) 492.
\bibitem{moya}
H. Garcilazo, E. Moya de Guerra, Nucl. Phys. {\bf A 562} (1993) 521.
\bibitem{gillespe}
J. Gillespie, {\it " Final State Interactions"}, Holden-Day Advanced Physics Monographs, 1964.
\bibitem{ankeExper}
S. Barsov et al., nucl-ex/0305031, Eur. Phys. J. {\bf A} (2004) in print.
\bibitem{HADES}
P. Salabura et al. (HADES Collaboration), Acta Phys. Polon. {\bf B 35} (2004) 1119.
\bibitem{quad}
L.P. Kaptari, A.Yu. Umnikov, S.G. Bondarenko, K.Yu. Kazakov,
F.C. Khanna, B. K\"ampfer, Phys. Rev. {\bf C 54} (1996) 986.
\bibitem{ourphysrev}
L.P. Kaptari, B. K\"ampfer, S.M. Dorkin,  S.S. Semikh,
Phys. Rev. {\bf C 57} (1998) 1097; Phys. Lett. {\bf B 404 } (1997) 8.
\bibitem{Mosel} 
V. Shklyar, G. Penner, U. Mosel, nucl-th/0403064 and further references therein.
\bibitem{M.Lutz} 
M.F.M. Lutz, E.E. Kolomeitsev, Nucl. Phys. {\bf A 700} (2002) 193;\\
E.E. Kolomeitsev, M.F.M. Lutz, Phys. Lett. {\bf 585} (2004) 243.
\bibitem{Sasha} 
A.I. Titov, B. K\"ampfer, B.L. Reznik, Nucl. Phys. {\bf A 721} (2003) 583;\\
A.I. Titov, B. K\"ampfer, Eur. Phys. J. {A 12} (2001) 217.
\bibitem{Fuchs2} 
C. Fuchs, M.I. Krivoruchenko, H.L. Yadav, A. Faessler, B.V. Martemyanov,
K. Shekhter, Phys. Rev. {\bf 67} (2003) 025202.
\end{thebibliography}
\end{document}